\documentclass[showpacs,prl,aps,onecolumn,superscriptaddress,psfrag,amsmath,amssymb]{revtex4}
\usepackage{float}
\usepackage{graphicx}
\usepackage[normalem]{ulem}
\usepackage{subfigure}
\usepackage{amssymb}

\begin{document}

\title{Crystalline Electric Field Randomness in the Triangular Lattice Spin-Liquid YbMgGaO$_4$}

\author{Yuesheng Li}
\email{yuesheng.man.li@gmail.com}
\affiliation{Department of Physics,
Renmin University of China,
Beijing 100872, P. R. China}
\affiliation{Experimental Physics VI, Center for Electronic Correlations and Magnetism, University of Augsburg, 86159 Augsburg, Germany}

\author{Devashibhai Adroja}
\affiliation{ISIS Pulsed Neutron and Muon Source, STFC Rutherford Appleton Laboratory, Harwell Campus, Didcot, Oxfordshire, OX11 0QX, United Kingdom}
\affiliation{Highly Correlated Matter Research Group, Physics Department,University of Johannesburg, PO Box 524, Auckland Park 2006, South Africa}

\author{Robert I. Bewley}
\affiliation{ISIS Pulsed Neutron and Muon Source, STFC Rutherford Appleton Laboratory, Harwell Campus, Didcot, Oxfordshire, OX11 0QX, United Kingdom}

\author{David Voneshen}
\affiliation{ISIS Pulsed Neutron and Muon Source, STFC Rutherford Appleton Laboratory, Harwell Campus, Didcot, Oxfordshire, OX11 0QX, United Kingdom}

\author{Alexander A. Tsirlin}
\affiliation{Experimental Physics VI, Center for Electronic Correlations and Magnetism, University of Augsburg, 86159 Augsburg, Germany}

\author{Philipp Gegenwart}
\affiliation{Experimental Physics VI, Center for Electronic Correlations and Magnetism, University of Augsburg, 86159 Augsburg, Germany}

\author{Qingming Zhang}
\email{qmzhang@ruc.edu.cn}
\affiliation{Department of Physics,
Renmin University of China, Beijing 100872, P. R. China}
\affiliation{Collaborative Innovation Center of Advanced Microstructures, Nanjing 210093, P. R. China}

\date{\today}

\begin{abstract}
We apply moderate-high-energy inelastic neutron scattering (INS) measurements to investigate Yb$^{3+}$ crystalline electric field (CEF) levels in the triangular spin-liquid candidate YbMgGaO$_4$. Three CEF excitations from the ground-state Kramers doublet are centered at the energies $\hbar \omega$ = 39, 61, and 97\,meV in agreement with the effective \mbox{spin-1/2} $g$-factors and experimental heat capacity, but reveal sizable broadening. We argue that this broadening originates from the site mixing between Mg$^{2+}$ and Ga$^{3+}$ giving rise to a distribution of Yb--O distances and orientations and, thus, of CEF parameters that account for the peculiar energy profile of the CEF excitations. The CEF randomness gives rise to a distribution of the effective spin-1/2 $g$-factors and explains the unprecedented broadening of low-energy magnetic excitations in the fully polarized ferromagnetic phase of YbMgGaO$_4$, although a distribution of magnetic couplings due to the Mg/Ga disorder may be important as well.
\end{abstract}


\maketitle

\emph{Introduction.}---Quantum spin liquid (QSL) is a novel state of matter with zero entropy and without conventional symmetry breaking even at zero temperature. Such states were proposed to host 'spinons', exotic spin excitations with fractional quantum numbers~\mbox{\cite{balents2010spin,lee2008end,wen2004quantum}}. Although many candidate QSL materials with two-dimensional or three-dimensional interaction topologies on the triangular, kagome, and pyrochlore lattices were reported~\cite{shimizu2003spin,yamashita2008thermodynamic,yamashita2009thermal,itou2008quantum,itou2010instability,yamashita2011gapless,shores2005structurally,han2012fractionalized,fu2015evidence,PhysRevLett.109.037208,PhysRevLett.110.207208,li2014gapless,PhysRevLett.88.077204,PhysRevX.1.021002}, they typically suffer from magnetic or non-magnetic defects~\cite{lee2007quantum,freedman2010site,li2012structure,li2013transition,PhysRevB.94.024438}, spatial anisotropy~\cite{shimizu2003spin,itou2008quantum,li2014gapless}, antisymmetric Dzyaloshinsky-Moriya anisotropy~\cite{moriya1960new,zorko2008dzyaloshinsky,PhysRevLett.102.257201}, and (or) interlayer magnetic couplings~\cite{PhysRevLett.102.257201,doi2004structural,PhysRevLett.108.057205} that reduce or even completely release magnetic frustration~\mbox{\cite{PhysRevLett.102.257201,PhysRevLett.108.057205,PhysRevB.68.224416,PhysRevLett.98.207204,PhysRevB.79.214415}}.

Many of the aforementioned shortcomings can be remedied in a new triangular antiferromagnet YbMgGaO$_4$ that was recently reported by our group~\cite{li2015gapless,li2015rare,PhysRevLett.117.097201}. No spin freezing was detected down to at least 0.048\,K, which is about 3\% of the nearest-neighbor interaction $J_0\sim 1.5$\,K~\cite{PhysRevLett.117.097201}. Residual spin entropy is nearly zero at 0.06\,K, excluding any magnetic transitions at lower temperatures~\cite{li2015gapless}. Below 0.4\,K, thermodynamic properties evidence the putative QSL regime with temperature-independent magnetic susceptibility $\chi=\text{const}$~\cite{PhysRevLett.117.097201} and power-law behavior of the magnetic heat capacity, $C_m\sim T^{2/3}$~\cite{li2015gapless}, the observations that are consistent with theoretical predictions for the U(1) QSL ground state (GS) on the triangular lattice~\cite{motrunich2005variational,lee2005u,motrunich2006orbital}.

Very recently, two inelastic neutron scattering (INS) studies of YbMgGaO$_4$~\cite{shen2016spinon,paddison2016continuous} reported continuous excitations at transfer energies of $0.1-2.5$\,meV extending well above the energy scale of the magnetic coupling $J_0\sim 0.13$\,meV. These spectral features were identified as fractionalized excitations ('spinons') from the QSL GS~\cite{shen2016spinon}. Surprisingly, though, magnetic excitations remain very broad in both energy and wave-vector ($Q$) even in the almost fully polarized state at 7.8\,T, where only narrow spin-wave excitations of an ordered ferromagnet are expected~\cite{paddison2016continuous}. This persistent broadening of magnetic excitations may be related to a very inconspicuous intrinsic structural disorder that we uncover and quantify by INS measurements at high energies, where crystalline electric field (CEF) excitations of Yb$^{3+}$ ions can be probed.

In this Letter, we report a comprehensive investigation of the CEF excitations in YbMgGaO$_4$. They are observed at the energies of $\hbar \omega$ = 39, 61, and 97\,meV and show not only a pronounced broadening, but also a very peculiar energy profile with a shoulder around 87 meV on the side of the 97 meV excitation. These peculiarities are rationalized by considering the frozen Mg/Ga disorder that affects the local environment of Yb$^{3+}$ and, thus, the CEF parameters. Their randomness gives rise to a distribution of electronic $g$-factors and explains the broadening of low-energy magnetic excitations, thus rendering structural randomness an important ingredient of the QSL physics in YbMgGaO$_4$.

\emph{Experimental technique.}---Moderate-high-energy INS data and low-energy INS data were collected, respectively, on the MERLIN~\cite{bewley2006merlin} and LET~\cite{bewley2011let} spectrometers at the ISIS pulsed neutron facility, Rutherford Appleton Laboratory, U.K.~\cite{supple}. Several incident energies $E_i$ were used at MERLIN. In the following, we focus on the data obtained with $E_i=153.5$\,meV that provides the best trade-off between the resolution and energy coverage~\footnote{Note that the data obtained with $E_i=307$\,meV extend to much higher energies and reveal a weak excitation, presumably of CEF origin, at 139\,meV. This feature has only 20\% of the intensity of the 97\,meV excitation and is likely due to a multiple scattering excitation, see Supplemental material for details.} Our YbMgGaO$_4$ ($m_{\rm Yb}$ = 14.03 g) and LuMgGaO$_4$ ($m_{\rm Lu}$ = 6.36 g) powder samples for the MERLIN experiment were prepared using solid-state reactions~\cite{li2015gapless}. Single crystals of YbMgGaO$_4$ for the LET experiment were grown by the floating zone technique~\cite{li2015rare}.

\begin{figure}[t]
\begin{center}
\includegraphics[width=10cm,angle=0]{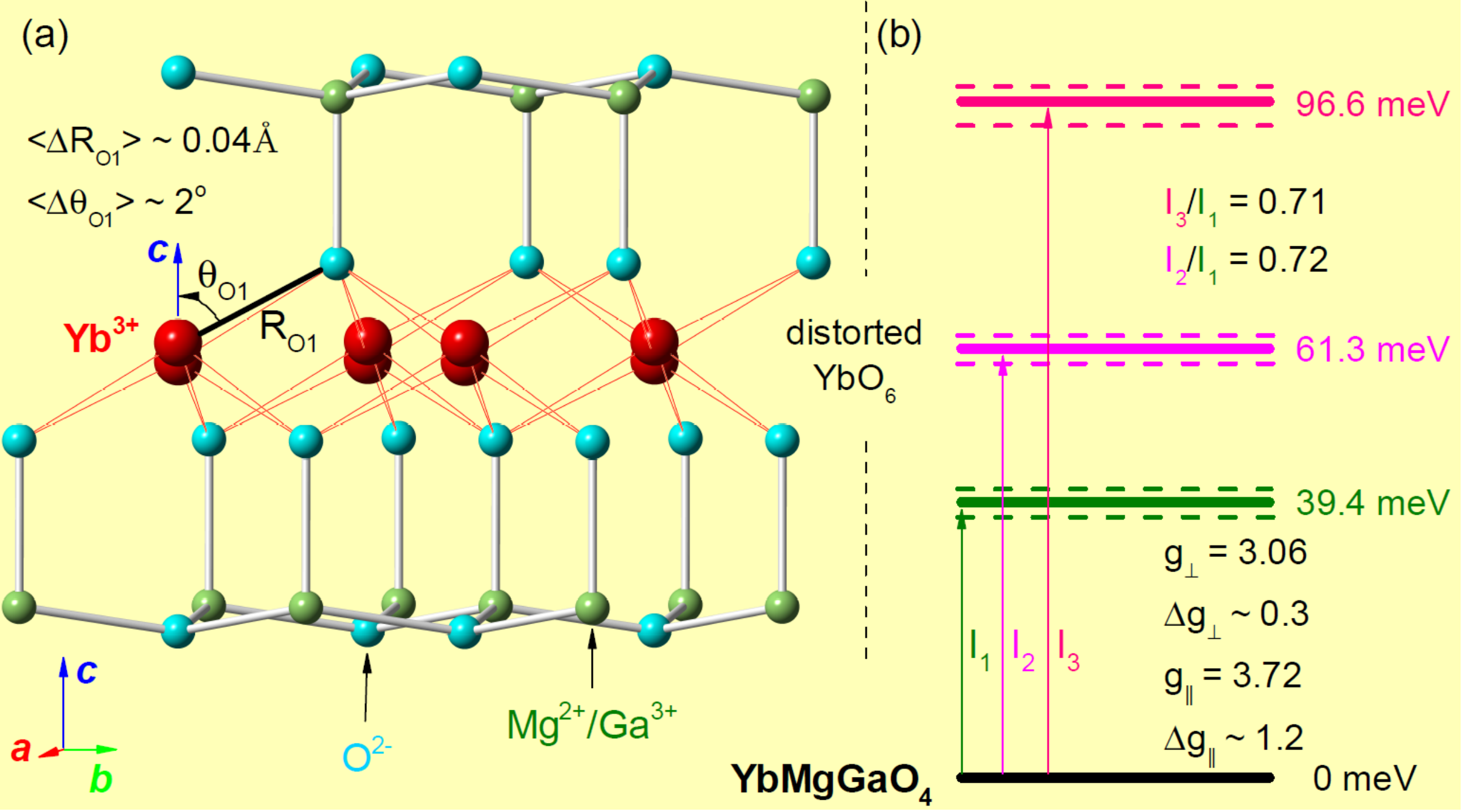}
\caption{(Color online)
(a) Crystal structure of YbMgGaO$_4$. The random distribution of Mg$^{2+}$ and Ga$^{3+}$ causes local distortions of the YbO$_6$ environments due to uneven charge distribution around the Yb$^{3+}$ site~\cite{supple}. (b) Four Kramers doublet energy levels and three CEF excitations obtained from the CEF fit. The dashed lines illustrate the broadening of the CEF excitations due to the inherent structural disorder.}
\label{fig1}
\end{center}
\end{figure}

\emph{CEF excitations.}---According to the Hund's rules, free Yb$^{3+}$ (4$f^{13}$) ion has the spin angular momentum $s=1/2$ and orbital angular momentum $L=3$ resulting in the eight-fold-degenerate ground state with the total angular momentum $J=7/2$ and Land\'{e} $g$-factor $g_J=8/7$ for the GS multiplet. In the idealized YbMgGaO$_4$ structure, Yb$^{3+}$ has trigonal local symmetry with the point group D$_{3d}$ (see Fig.~\ref{fig1}(a)) that splits this multiplet into four Kramers doublets~\cite{supple} (see Fig.~\ref{fig1}(b)).

\begin{figure}[t]
\begin{center}
\includegraphics[width=10cm,angle=0]{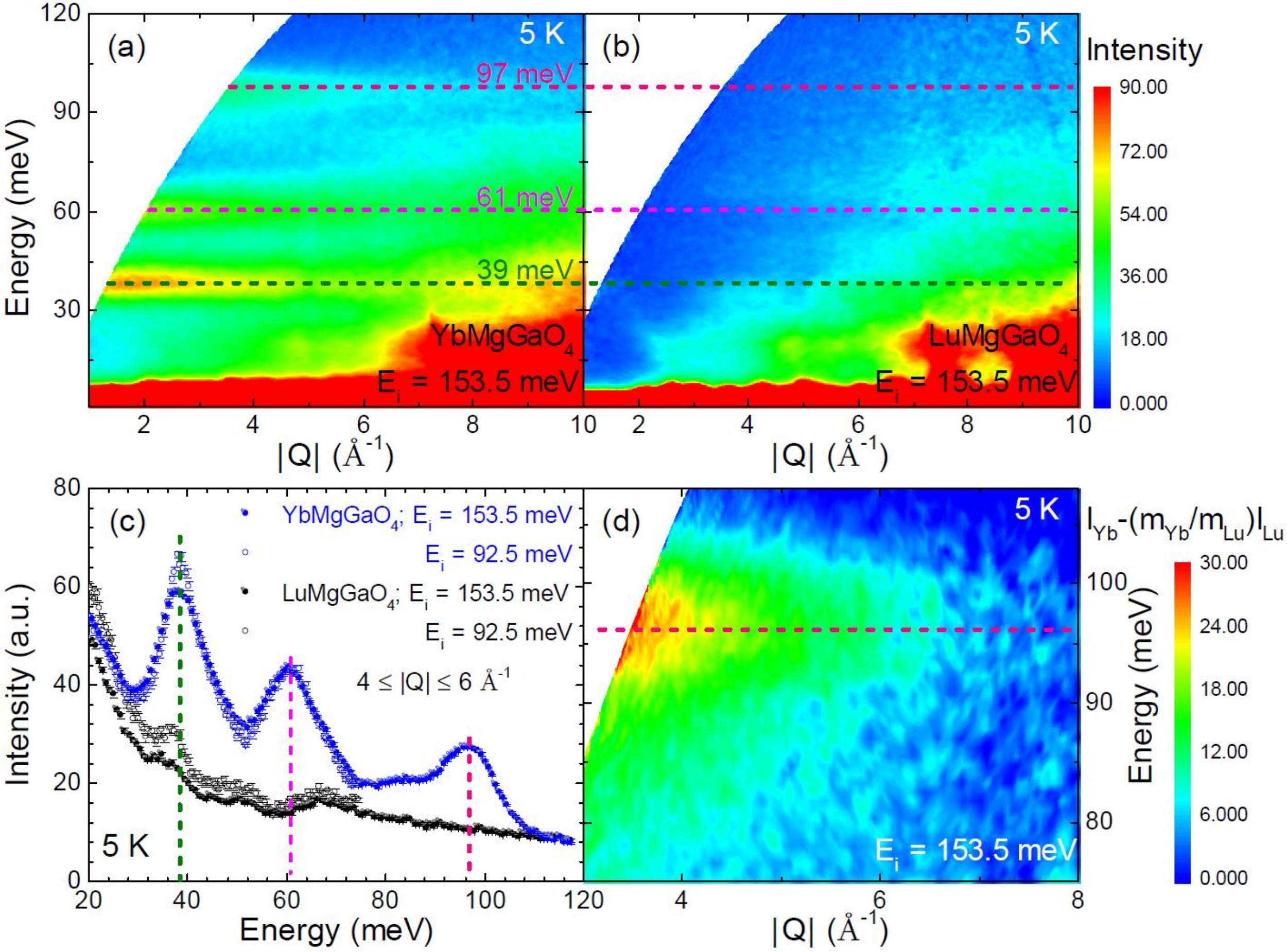}
\caption{(Color online)
MERLIN INS spectra for (a) YbMgGaO$_4$ and (b) LuMgGaO$_4$ measured with the incident neutron energy of 153.5\,meV at 5\,K. (c) Energy dependence of the INS intensity at 5\,K for both YbMgGaO$_4$ and LuMgGaO$_4$ measured with different incident neutron energies. The data have been integrated over the wave-vector space, $4\leq|Q|\leq 6$\,\r A$^{-1}$. Three CEF excitations of Yb$^{3+}$ are highlighted by colored dashed lines. The INS intensities of LuMgGaO$_4$ are multiplied by $m_{\rm Yb}$/$m_{\rm Lu}$. (d) Measured CEF excitation around 97 meV in a zoom-in view.}
\label{fig2}
\end{center}
\end{figure}

\begin{figure}[t]
\begin{center}
\includegraphics[width=10cm,angle=0]{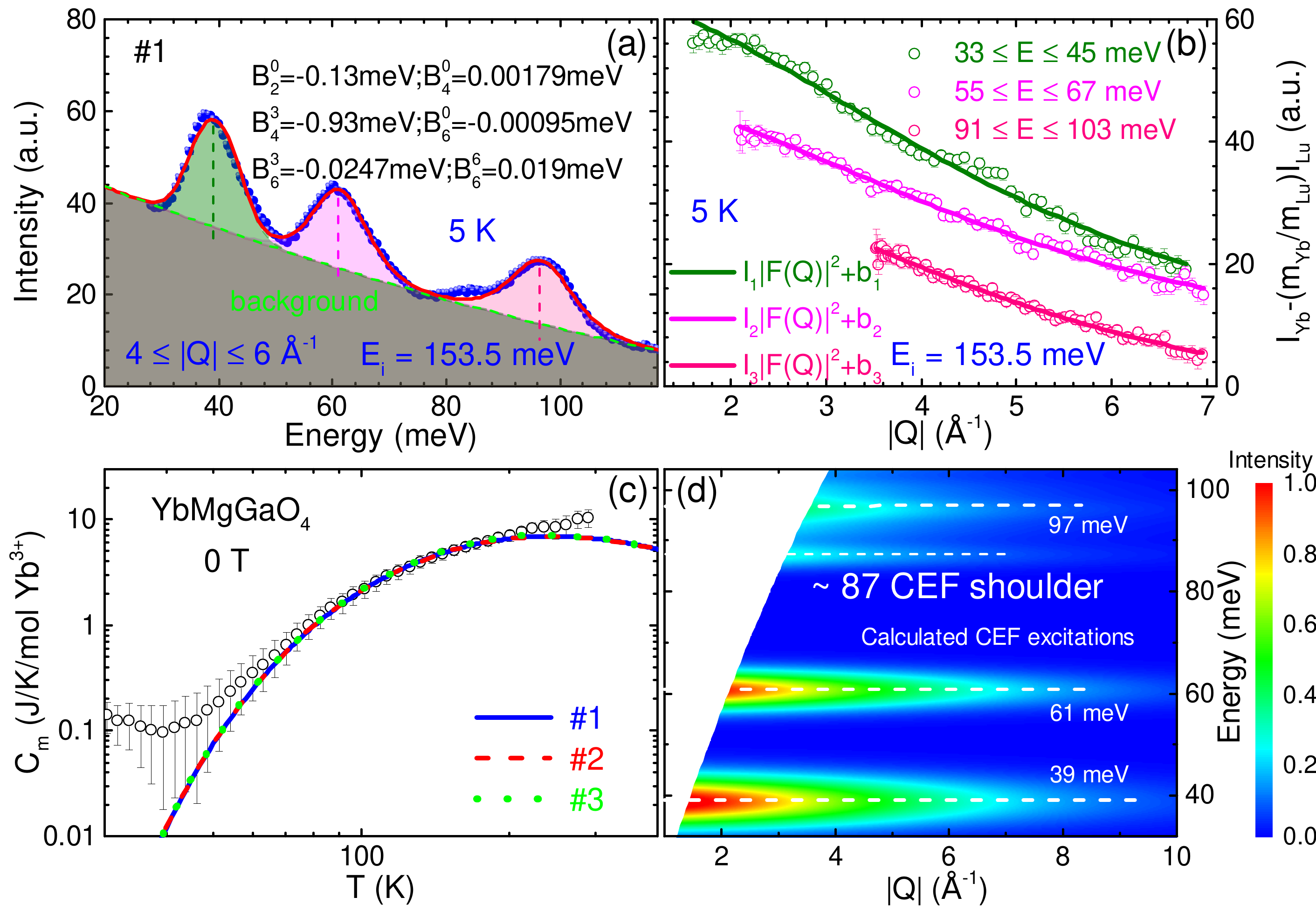}
\caption{(Color online)
(a) Peak fit to the INS spectra of YbMgGaO$_4$ (\#1) at 5 K. Three peak centers are obtained (colored dashed lines). (b) Wave-vector ($|Q|$) dependence of the INS intensities around the three CEF excitations. Lattice contributions are subtracted by the nonmagnetic counterpart ($(m_{\rm Yb}/m_{\rm Lu})\cdot I_{\rm Lu}$) measured on LuMgGaO$_4$. Colored curves show the fit with a small constant background (i.e. $I_k|F(Q)|^2+b_k$) to the integrated INS data, where $I_k$ $\gg$ $|b_k|$. (c) Temperature dependence of the magnetic heat capacity measured on YbMgGaO$_4$ single crystals. Lattice contribution is subtracted using the heat capacity of LuMgGaO$_4$~\cite{li2015rare}. The blue solid, red dashed and green dotted curves show the calculated heat capacities using three series of the fitted CEF parameters (\#1, \#2 and \#3)~\cite{supple} respectively. (d) Calculated INS spectra by considering different nearest-neighbor oxygen environments (distorted YbO$_6$ octahedrons)~\cite{supple} convoluted with the corresponding instrumental resolutions.}
\label{fig3}
\end{center}
\end{figure}

Raw INS spectra for both YbMgGaO$_4$ and its nonmagnetic analog LuMgGaO$_4$ are shown in panels (a) and (b) of Fig.~\ref{fig2}, respectively. Their comparison reveals three features that are identified as CEF excitations of Yb$^{3+}$ based on the following observations: (1) These excitations are absent in the non-magnetic reference compound LuMgGaO$_4$ (Fig.~\ref{fig2} (b)). (2) The lowest-lying excitation at around $\sim$ 39.4\,meV (see Fig.~\ref{fig2} (c)) is consistent with the energy separation $\Delta$ $\sim$ 36.5(1)\,meV between the ground-state Kramers doublet and the first excited state, as found in our previous heat capacity measurements~\cite{li2015rare}. (3) No systematic anharmonic effect is observed, thus excluding phonon origin of the excitations~\cite{supple,PhysRevB.28.1928}. (4) $Q$-independent excitation energies (see Fig.~\ref{fig2} (a)) suggest their single-ion nature. (5) The intensities decrease with $Q$ following the square of the magnetic form factor of the Yb$^{3+}$ ion  (see Fig.~\ref{fig3} (b)). (6) The lowest excitation at around 39\,meV is far above $J_0\simeq 0.13$\,meV and, therefore, unrelated to the spin-spin correlations in YbMgGaO$_4$. All these facts indicate that three spectral features are single-ion CEF excitations. For an Yb$^{3+}$ ion with $J=7/2$ in the D$_{3d}$ symmetry, we indeed expect four CEF doublets and thus three CEF excitations from the ground-state doublet.

A closer inspection of these CEF excitations reveals two unusual features, though. First, all excitations are much broader than the instrumental resolution. For example, at $E_i$ = 153.5\,meV the total FWHMs (full width at half maximum obtained from the convoluted Lorentzian and Gaussian peak profiles) are 10.1(4)\,meV ($\hbar\omega_1\sim39$\,meV), 10.9(4)\,meV ($\hbar\omega_2\sim61$\,meV), and 12.2(7)\,meV ($\hbar\omega_3\sim97$\,meV), much larger than the instrumental resolutions (Gaussian component) of 6.7, 5.6, and 4.3\,meV, respectively. Through convolution calculations, we determine the additional broadening (Lorentzian component) of 5.5, 7.9, and 10.7 meV, respectively. Given the high quality of our sample~\cite{supple} and the low temperature of the measurement ($T=5$\,K), we conclude that this broadening is intrinsic.

Another peculiar feature is the shape of the highest CEF excitation that shows the main peak around 97 meV and a shoulder at $\sim$ 87 meV (see Fig.~\ref{fig2} (d)). The $Q$-dependence of the intensity at $\sim$ 87\,meV follows the square of the magnetic form factor of the Yb$^{3+}$ ion~\cite{supple}, thus proving the CEF origin of this spectral feature. It contributes about 40\% of the overall intensity of the highest-energy excitation and is clearly intrinsic. Further, there are no phonon modes observed between $\sim$ 70 and 120 meV in LuMgGaO$_4$ (see Fig. ~\ref{fig2}(b)) and hence the $\sim$ 87 meV shoulder could not be due to CEF-phonon coupling~\cite{PhysRevLett.108.216402}.

\emph{Combined CEF fit.}---To determine the CEF parameters, we fit energy dependence of the experimental intensity in four fashions (see Fig.~\ref{fig3} (a) and~\cite{supple}). Fit \#1 is performed against the whole dataset with a three-peaks fit (see Fig.~\ref{fig3} (a)), for fit \#2 we excluded the region between 73 and 90\,meV, where the additional shoulder is observed, while fit \#3 is performed against the whole dataset with a four-peaks fit and uses the additional shoulder energy ($\sim$ 87 meV) as $\hbar\omega_3$, and fit \#4 uses the additional mode energy (138.7(3) meV) as $\hbar\omega_3$. All fits share the same measured relative INS intensities, $I_2/I_1$ and $I_3/I_1$
(see Fig.~\ref{fig3} (b)), and the same measured effective spin-
1/2 g-factors~\cite{li2015rare}. Through combined fits to these seven observables -- $\hbar \omega_1$, $\hbar \omega_2$, $\hbar \omega_3$, $I_2/I_1$, $I_3/I_1$, $g_{\perp}$, and $g_{\parallel}$ -- we obtain all six CEF parameters, $B_n^m$~\cite{supple}, by minimizing the following deviation function,
\begin{equation}
R_p=\sqrt{\frac{1}{7}\sum_{i=1}^{7}\left(\frac{X_i^{\rm obs}-X_i^{\rm cal}}{\sigma_i^{\rm obs}}\right)^2},
\label{eq1}
\end{equation}
where $X_i^{\rm obs}$ and $\sigma_i^{\rm obs}$ are the experimental value and its standard deviation, respectively, whereas $X_i^{\rm cal}$ is the calculated value. Qualitatively similar CEF parameters and wavefunctions are found from fits \#1, \#2,  \#3 and \#4, as shown in~\cite{supple}, respectively. Magnetic part of the experimental heat capacity ($C_m$) is very well reproduced with the first three parameter sets (see Fig. ~\ref{fig3} (c)):
\begin{equation}
C_m^{\rm CEF}=\frac{1}{k_{B}T^{2}}\frac{\partial^2\ln[\sum_{k=0}^{3}2 \exp(-\frac{\hbar\omega_k}{k_{B}T})]}{\partial(\frac{1}{k_{B}T})^2}.
\label{eq2}
\end{equation}

\emph{Inherent structural disorder and CEF randomness.}---Peculiar shape of the CEF excitations is rooted in subtle details of the YbMgGaO$_4$ crystal structure. Our single-crystal x-ray diffraction study excludes any global symmetry reduction or a site mixing between Yb and Mg/Ga~\cite{supple}. On the other hand, Mg and Ga share one crystallographic site, thus forming different local configurations around each Yb$^{3+}$ ion. The most obvious effect of this Mg/Ga disorder is the variation of the electrostatic potential imposed on Yb$^{3+}$. We assess it by calculating CEF parameters using the point-charge model~\cite{supple} and find that, as long as Yb occupies its ideal position at (0,0,0), the random distribution of Mg and Ga gives rise to only a weak broadening of the CEF excitations, $\Delta(\hbar\omega_1)$ = 0.27 meV, $\Delta(\hbar\omega_2)$ = 0.26 meV, and $\Delta(\hbar\omega_3)$ = 0.39 meV, and does not account for our experimental observations. Moreover, all three CEF excitations remain symmetric.

A further effect of the Mg/Ga disorder is local charge misbalance that may push Yb out of its ideal position, as reflected by the enhanced values of the Yb atomic displacement parameter, with the thermal ellipsoid elongated along the $c$-direction~\cite{supple,cava1998compounds}. We probed this effect quantitatively by constructing several representative Mg/Ga configurations and optimizing their geometry using density-functional calculations~\cite{supple}. We indeed observed that exact positions of both Yb and its neighboring oxygens are affected by the local distribution of Mg$^{2+}$ and Ga$^{3+}$. The resulting distortions of the YbO$_6$ octahedra give rise to a pronounced distribution of the CEF parameters and render the highest-lying CEF excitation asymmetric~\cite{supple}. Both the $\sim$ 87 meV shoulder and the overall broadening of the CEF excitations can be well reproduced (see Fig.~\ref{fig3} (d) and~\cite{supple}). Consequently, the effective spin-1/2 $g$-factors show a maximum distribution as follows: $\Delta g_{\perp}$ $\sim$ 0.3 and $\Delta g_{\parallel}$ $\sim$ 1.2~\cite{supple}. This distribution of the $g$-values has immediate ramifications for low-energy excitations, as we show below. Eventually, intersite magnetic couplings should be random too~\cite{PhysRevB.86.174424}, although in a much more complicated manner.

Prior to discussing the low-energy excitations, let us note that the broadening of CEF excitations is not uncommon for rare-earth compounds~\cite{PhysRevB.91.224430,PhysRevB.92.144422,wen2016disordered}. This effect is usually ascribed to antisite defects, as in the ``stuffed'' quantum spin ice, Yb$_2$Ti$_{2-x}$Yb$_x$O$_{7-x/2}$ with $x\sim 0.01-0.02$~\cite{PhysRevB.86.174424,PhysRevB.92.134420,PhysRevB.93.064406}. On the other hand, site mixing beyond the rare-earth site, as in Tb$_2$Sn$_{2-x}$Ti$_x$O$_7$, is believed to merge the CEF excitations into a broad continuum~\cite{PhysRevB.91.245141}. Interestingly, YbMgGaO$_4$ with its complete Mg/Ga disorder and \textit{without} any detectable Yb antisite defects retains well-defined CEF excitations, albeit with a peculiar energy profile that can be reproduced, perhaps for the first time, by considering local atomic relaxation depending on the distribution of Mg and Ga around Yb$^{3+}$.

\begin{figure}[t]
\begin{center}
\includegraphics[width=10cm,angle=0]{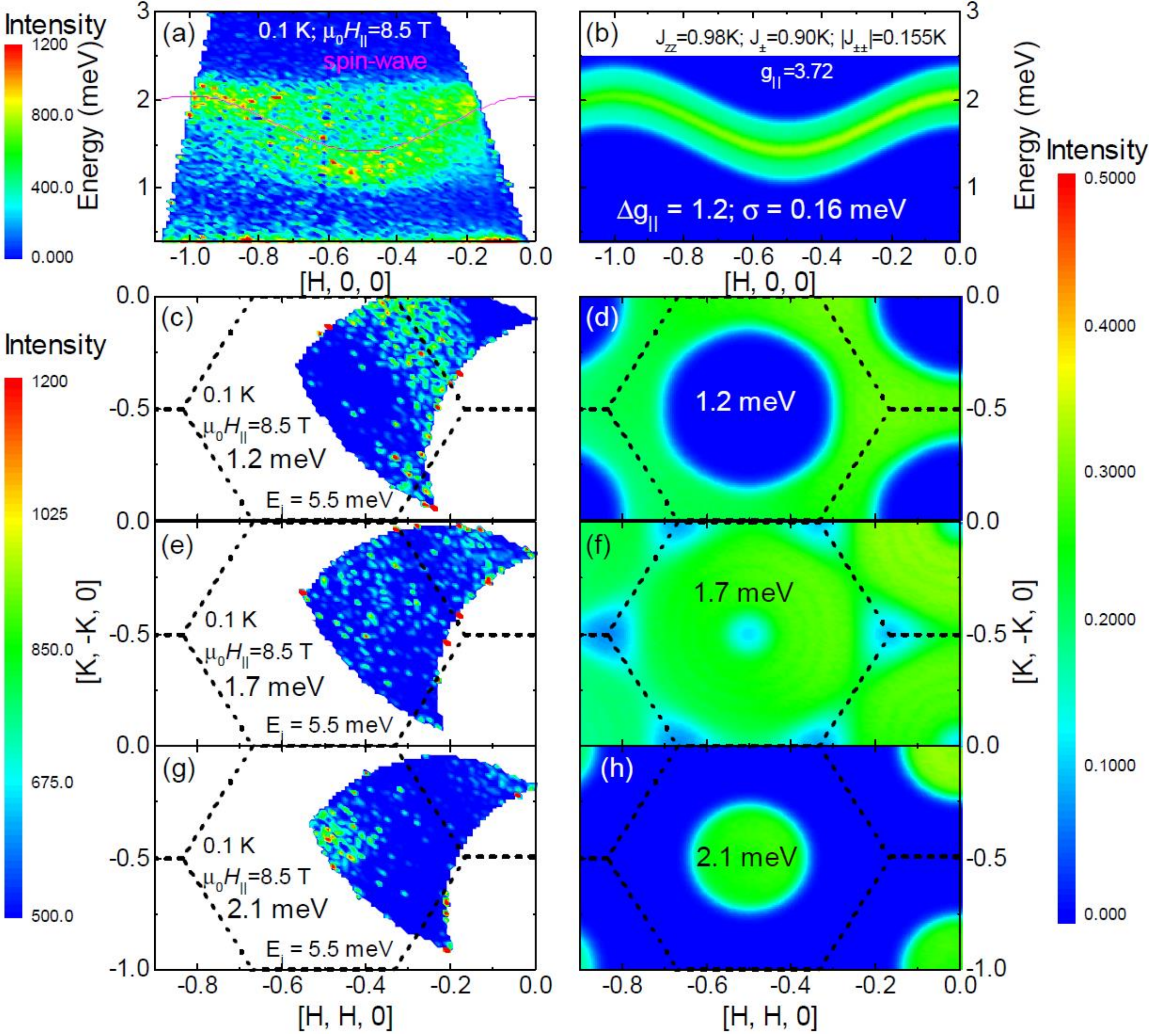}
\caption{(Color online)
LET INS spectra of a YbMgGaO$_4$ single crystal sample measured at 0.1 K under a field of 8.5 T applied along the \emph{c}-axis, the incident neutron energy is 5.5 meV. (a) Energy dependence of the excitations along the wave-vector direction [H, 0, 0]. (c), (e) and (g) Wave-vector dependence of the excitations at the transfer energies 1.2, 1.7 and 2.1 meV, respectively. (b), (d), (f) and (h) Calculated spin-wave excitations using the previously reported effective spin-1/2 g-factor and coupling constants~\cite{li2015rare} by considering a convolution of a broadening of $g_{\parallel}$ ($\Delta g_{\parallel}$ = 1.2) and a instrumental Gaussian broadening ($\sigma$ = 0.16 meV). The calculated spin-wave dispersion without any broadening is shown in (a) (pink line). The black dashed lines represent Brillouin zone boundaries.}
\label{fig4}
\end{center}
\end{figure}

\emph{Spin-wave continuum.}---Under the field of 8.5 T applied along the \emph{c}-axis, the spin system is fully polarized according to the static magnetization measurement at 1.9 K~\cite{supple}. Therefore, at 8.5 T and 0.1 K the fully polarized (ferromagnetic) state should give rise to narrow spin-wave excitations having the width of about 0.16 meV according to the instrumental resolution of LET. In contrast, the measured INS signals are still broadly distributed in the energy space with a width of more than 0.5 meV (see Fig.~\ref{fig4} (a)). This width is obviously larger than the instrumental resolution~\cite{bewley2011let} and than the width of spin-wave excitations in single crystals of a similar Yb$^{3+}$ material, Yb$_2$Ti$_2$O$_7$~\cite{PhysRevX.1.021002}. Similarly broad excitations were observed in the recent INS measurement performed in the applied field of 7.8 T at 0.06 K~\cite{paddison2016continuous}. Our magnetization data~\cite{supple} confirm that at 8.5\,T the Yb$^{3+}$ spins are fully polarized along the \emph{c}-axis. Nevertheless, the excitations remain very broad. Thus, the natural explanation to this observed spin-wave continuum is the aforementioned randomness of the effective spin-1/2 $g$-factors and (or) couplings, since linear spin-wave theory should be applicable to YbMgGaO$_4$ at 8.5 T and 0.1 K.

We model the spin-wave excitations in the \emph{ab}-plane under the high applied field along the \emph{c}-axis using the expression~\cite{xu2013absolute}
\begin{equation}
\frac{d^2\sigma}{d\Omega d\omega} \propto |F(\mid\textbf{Q}\mid)|^2\times[\frac{Q_y^2S^{xx}(\textbf{Q},E)+Q_x^2S^{yy}(\textbf{Q},E)}{\mid\textbf{Q}\mid^2}+S^{zz}(\textbf{Q},E)],
\label{eq3}
\end{equation}
where $F(\mid\textbf{Q}\mid)$ is the magnetic form factor of Yb$^{3+}$ and $S^{\alpha\alpha}(\textbf{Q},E)$ ($\alpha$ = $x$, $y$ or $z$) is the dynamic spin structure factor calculated by the Spinw-Matlab code based on the linear spin-wave theory~\cite{toth2015linear} and coupling parameters reported earlier~\cite{li2015rare}. By considering the maximum broadening of $g_{\parallel}$, $\Delta g_{\parallel}$ = 1.2, and the instrumental Gaussian broadening, $\sigma$ = 0.16 meV, we are able to reproduce the broadening of about 0.7\,meV at the zone center. However, the signal is even broader at the zone boundary. While the on-site randomness is clearly very important for the low-energy physics, a distribution of magnetic couplings is relevant too and requires further investigation.

\emph{Conclusions.}---The CEF excitations of Yb$^{3+}$ in the triangular QSL YbMgGaO$_4$ have been studied by moderate-high-energy INS measurements. Large broadening and peculiar energy profile of the CEF excitations is observed and ascribed to the structural randomness, namely, the random distribution of Mg and Ga that affects local coordination of the Yb$^{3+}$ ions. We propose that this inherent structural disorder results in the distribution of the effective spin-1/2 $g$-factors that is responsible for the persistent broadening of low-energy magnetic excitations in the fully polarized ferromagnetic state, although the distribution of magnetic couplings seems to be relevant too. Our results put forward structural randomness as an important ingredient in the spin-liquid physics, an observation that goes hand in hand with the recent report on the suppression of thermal conductivity at low temperatures~\cite{PhysRevLett.117.267202}. In 4$f$-based materials, the randomness of CEF levels can be easily introduced without generating strong structural disorder, thus opening interesting prospects for the design of new spin-liquid materials~\cite{sanders2016magnetism}.

\emph{Note added.}---After the submission of our paper, the results of Ref.~\onlinecite{paddison2016continuous} were updated as it progressed from a preprint to a published article incorporating high-energy INS data that are largely consistent with our results.

\emph{Acknowledgment}---We thank Jinchen Wang and Yu Li for helpful discussion. YL thanks Duc Le for introducing him to the Mantid (3.7.1) software for CEF fits. AT acknowledges Tom Fennell for pointing our recent examples of CEF excitations broadening. This work was supported by the NSF of China and the Ministry of Science and Technology of China (973 projects: 2016YFA0300504). Y.S.L. was supported by the start-up funds of Renmin University of China. The work in Augsburg was supported by the German Federal Ministry for Education and Research through the Sofja Kovalevskaya Award of Alexander von Humboldt Foundation. Q.M.Z. was supported by the Fundamental Research Funds for the Central Universities, and by the Research Funds of Renmin University of China.
\\
\\
\\
\\
\\
\\
\textit{Supplementary material}\medskip \\ Crystalline Electric Field Randomness in the Triangular Lattice Spin-Liquid YbMgGaO$_4$
\\
\\
\\
We present here:

1. General techniques.

   ~ ~1.1 Moderate-high-energy MERLIN inelastic neutron scattering (INS) experiments.

   ~ ~1.2 Low-energy LET INS experiments.

   ~ ~1.3 Crystalline electric field (CEF) Hamiltonian and its relationship to observations.

   ~ ~1.4 CEF fits.

   ~ ~1.5 Calculation of the CEF parameters using the point-charge model.

2. X-ray diffraction (XRD) patterns for YbMgGaO$_4$ and LuMgGaO$_4$ powders used in the MERLIN INS experiments.

3. First-derivative absorption electron spin resonance (ESR) spectra for (Yb$_{0.04}$Lu$_{0.96}$)MgGaO$_4$ and YbMgGaO$_4$~\cite{li2015gapless}.

4. Shape of the 97 meV CEF excitation.

5. Energy dependence of the CEF excitations measured for YbMgGaO$_4$ from 5 to 300 K.

6. Combined CEF fit results for YbMgGaO$_4$.

7. Fitted CEF wavefunctions.

8. Crystal structure of YbMgGaO$_4$ at 100 K determined from single-crystal XRD data.

9. Local structural configurations of Yb$^{3+}$ obtained from density-functional band-structure calculations.

10. Calculated CEF randomness in the framework of the point-charge model.

   ~ ~10.1 Calculated CEF randomness caused by the distribution of nearest-neighbor oxygen environments.

   ~ ~10.2 Calculated CEF randomness caused solely by the site-mixing between Mg$^{2+}$ and Ga$^{3+}$.

11. INS spectra measured at $E_i$ = 307\,meV, and the 138.7(3) meV mode.

12. Elastic neutron diffraction pattern for the single crystal sample of YbMgGaO$_4$ in the \emph{ab}-plane used in LET INS experiments.

13. Magnetic field dependence of the susceptibility ($dM/dH$) measured at 1.9 K along the \emph{c}-axis.

\section{1. general techniques}

\subsection{1.1 Moderate-high-energy Merlin INS experiments}

The MERLIN measurements between 5 and 300\,K were performed on both YbMgGaO$_4$ and LuMgGaO$_4$ samples with the incident beam energies $E_i$ = 92.5, 153.5, and 307\,meV. At $E_i$ = 92.5\,meV, the instrumental resolutions (full width at half maximum, FWHM) are 2.9 and 2.3\,meV at the transfer-energies of 39 and 61\,meV, respectively, whereas at $E_i$ = 153.5\,meV, the resolutions are 6.7, 5.6, and 4.3\,meV at 39, 61, and 97\,meV, respectively~\cite{bewley2006merlin}. And at $E_i$ = 307\,meV the instrumental resolution is 22.5\,meV at $\hbar\omega$ $\sim$ 0 meV. Phase purity of both samples was confirmed by X-ray diffraction prior to the MERLIN INS measurements (see Fig.~\ref{figs1}). All MERLIN data were processed and cut using the Mslice-Matlab code.

\subsection{1.2 Low-energy LET INS experiments}

Incident energies of 26.8 and 5.5 meV were chosen for both elastic and inelastic scattering with the energy resolution of 1.4 and 0.16 meV, respectively~\cite{bewley2011let}. The as-grown rod ($\sim$ 50 g) was cut into slices along the \emph{ab}-plane. Ten best-quality \emph{ab}-slices of the single-crystal (total mass $\sim$ 10 g) were selected for the neutron scattering experiment on LET by Laue X-ray diffraction on all surfaces. The slices were fixed to the copper base by CYTOP glue to avoid any shift in applied magnetic fields up to 8.5 T. The sample temperature of 0.1 K was achieved using a dilution refrigerator. The elastic neutron diffraction (see Fig.~\ref{figs8}) showed that the alignment of the single crystals was sufficient for the INS study of the continuous excitations.

All INS data shown in the main text have been integrated over the momentum space, $-0.9\leq\eta\leq 0.9$ in $[0,0,-\eta]$. In Fig. 4 (a) of the main text, the INS data have been also integrated over $-0.03\leq\xi\leq 0.03$ in $[\xi/2,-\xi,0]$. In Fig. 4 (c), (e) and (g) of the main text, the INS data have been integrated over small energy ranges, (c) 1.15 $\leq$ $E$ $\leq$ 1.25 meV, (e) 1.65 $\leq$ $E$ $\leq$ 1.75 meV and (g) 2.05 $\leq$ $E$ $\leq$ 2.15 meV, respectively. All LET data were processed and cut using the Horace-Matlab on the ISIS computers.

\subsection{1.3 Crystalline electric field (CEF) Hamiltonian and its relationship to observations}

Under zero applied magnetic field, the generic CEF Hamiltonian that is invariant under the D$_{3d}$ point group symmetry is given by~\cite{bertin2012crystal,PhysRevB.76.184436,PhysRevB.94.024430}
\begin{equation}
H_0=B_{2}^{0}O_{2}^{0}+B_{4}^{0}O_{4}^{0}+B_{4}^{3}O_{4}^{3}+B_{6}^{0}O_{6}^{0}+B_{6}^{3}O_{6}^{3}+B_{6}^{6}O_{6}^{6},
\label{eqs1}
\end{equation}
where $B_{n}^{m}$ ($n,m$ are integers and $n\geq m$) are CEF parameters that will be determined experimentally, and the Stevens operators $O_{n}^{m}$ are polynomial functions of the components of the total angular momentum operator $J_z$, $J_+$, and $J_-$ ($J_{\pm}=J_x\pm iJ_y$). For the moderate-high-energy INS measurements ($>10$\,meV), it is sufficient to consider only the single-ion CEF Hamiltonian [Eq.~\eqref{eqs1}] and neglect the couplings between the Yb$^{3+}$ ions ($J_0\sim 0.13$\,meV)~\cite{li2015rare}. Under the eight-dimensional representation space $|J=7/2,m_J\rangle$ ($m_J$ = $\pm$1/2, $\pm$3/2, $\pm$5/2 and $\pm$7/2), we diagonalize the single-ion CEF Hamiltonian and obtain four ascending energy levels, $\hbar \omega_{k}$ (k = 0, 1, 2 and 3). Each energy level is two-fold-degenerate with the eigenfunctions (Kramers doublet)
\begin{equation}
 |\omega_{k,\pm}\rangle = \sum_{m_J = -7/2}^{7/2}C_{m_J}^{k,\pm}\, |J=7/2, m_J\rangle,
\label{eqn}
\end{equation}
where $|C_{m_J}^{k,\pm}|=|C_{-m_J}^{k,\mp}|$.

In a paramagnetic system (single-ion approximation), the differential neutron cross section of the powder sample for the CEF Kramers doublet transition $|\omega_{k1}\rangle \rightarrow |\omega_{k2}\rangle$ is given in the dipole approximation by~\cite{bertin2012crystal,PhysRevB.76.184436,PhysRevB.94.024430,furrer1988neutron,PhysRevB.78.144422,PhysRevLett.101.217002,PhysRevB.92.144422}:
\begin{equation}
\frac{d^2\sigma}{d\Omega d\omega} \propto |F(Q)|^2 \exp\left(\frac{-\hbar\omega_{k1}}{k_B T}\right)\times\sum_{\alpha=x,y,z}|\langle\omega_{k2}|J_\alpha|\omega_{k1}\rangle|^2\delta(\omega_{k1}-\omega_{k2}+\omega),
\label{eqs2}
\end{equation}
where $F(Q)$ = $\langle j_0(\frac{Q}{4\pi})\rangle$ + $\frac{2-g_J}{g_J}\langle j_2(\frac{Q}{4\pi})\rangle$ is the magnetic form factor in the dipole approximation, and $|\langle\omega_{k2}|J_{\alpha}|\omega_{k1}\rangle|^2 = |\langle\omega_{k2,+}|J_{\alpha}|\omega_{k1,+}\rangle|^2 + |\langle\omega_{k2,-}|J_{\alpha}|\omega_{k1,+}\rangle|^2 + |\langle\omega_{k2,+}|J_{\alpha}|\omega_{k1,-}\rangle|^2 + |\langle\omega_{k2,-}|J_{\alpha}|\omega_{k1,-}\rangle|^2$ is proportional to the total transition probability. At 5 K, only three CEF excitations from the ground Kramers doublet are taken into account, and the CEF transitions from the excited doublets are ignored.

At low temperatures ($T \ll \hbar(\omega_{1}-\omega_{0})/k_B$), the effective spin-1/2 $g$-factors for the ground-state Kramers doublet ($|\omega_{0,\pm}\rangle$) are given by:
\begin{gather}
g_{\perp}=g_J|\langle\omega_{0,\pm}|J_{\pm}|\omega_{0,\mp}\rangle|
\label{eqs3} \\
g_{\parallel}=2g_J|\langle\omega_{0,\pm}|J_z|\omega_{0,\pm}\rangle|
\label{eqs4}
\end{gather}
for the field perpendicular and parallel to the $c$-axis, respectively

\subsection{1.4 CEF fits}

All CEF simulations and combined fits to the experimental data (excitation energies, relative INS intensities, and effective spin-1/2 g-factors) were performed using Matlab programs. Our Matlab code for the CEF calculations (excitation energies and relative INS intensities) is fully compatible with the widely used code \texttt{Mantid} (3.7.1).

\subsection{1.5 Calculation of the CEF parameters using the point-charge model}

CEF parameters can be calculated for a given crystal structure within the point-charge approximation using the expression~\cite{baldovi2013simpre}
\begin{equation}
B_n^m = C_{\rm Yb}(n,m)\sum_{i} \frac{Z_{i}Y_n^m (\theta_i,\varphi_i)}{R_i^{n+1}},
\label{eqs5}
\end{equation}
where $Z_{i}$ is the net charge (unit: $e$) of the lattice ion $i$ with the spherical coordinates ($R_i$,$\theta_i$,$\varphi_i$), $Y_n^m$ are spherical harmonics, and $C_{\rm Yb}(n,m)$ is a pre-factor that depends on $n$ and $m$ only. Since $B_n^m \propto \frac{1}{R_i^{n+1}}$ ($n$ $\geq$ 2), ions beyond nearest neighbors can be neglected. On the other hand, the point-charge calculation does not take into account effects of exchange fields and may not deliver exact values of $B_n^m$. Here, we used the point-charge calculations to obtain the distribution of individual CEF parameters for different scenarios of structural disorder, and eventually re-scaled average values of these distributions to the experimental (fit \#1) CEF parameters listed in Table~\ref{table1}.

\section{2. XRD patterns for YbMgGaO$_4$ and LuMgGaO$_4$ powders used in the MERLIN INS experiments}

\begin{figure}[H]
\centering
\includegraphics[width=8.2cm,angle=0]{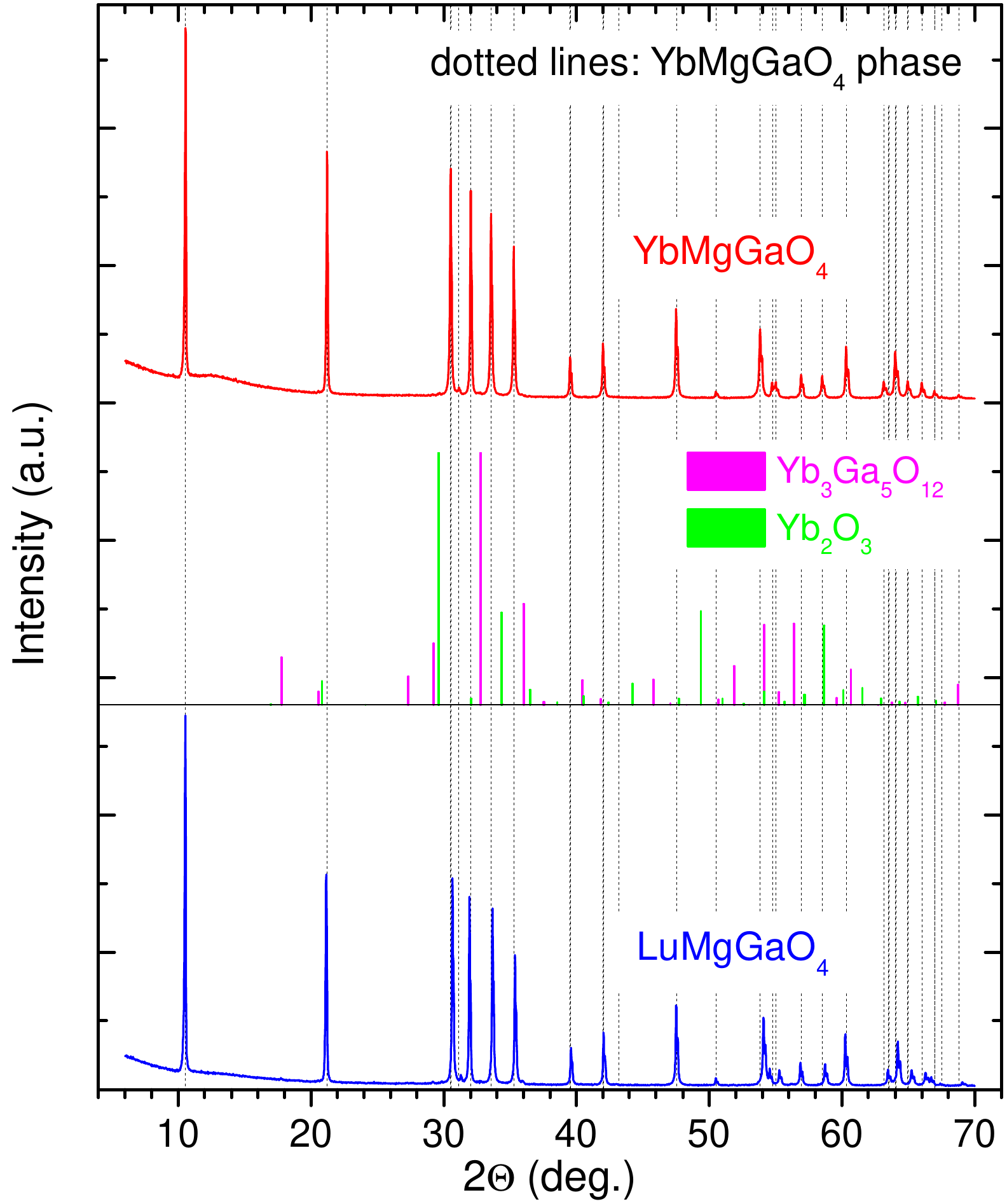}
\caption{(Color online.)
Powder XRD patterns for YbMgGaO$_4$ and LuMgGaO$_4$. No Yb$^{3+}$-related impurities, such as Yb$_3$Ga$_5$O$_{12}$ and Yb$_2$O$_3$, are observed.}
\label{figs1}
\end{figure}

Fig.~\ref{figs1} shows the XRD patterns for the YbMgGaO$_4$ and LuMgGaO$_4$ powder samples used in the MERLIN INS measurements. No reflections of foreign phases are observed, suggesting that the concentration of possible Yb$^{3+}$-related impurities, such as Yb$_3$Ga$_5$O$_{12}$ and Yb$_2$O$_3$, should be well below $\sim$ 5\%.

\section{3. First-derivative absorption ESR spectra for (Yb$_{0.04}$Lu$_{0.96}$)MgGaO$_4$ and YbMgGaO$_4$~\cite{li2015gapless}}

\begin{figure}[H]
\centering
\includegraphics[width=10cm,angle=0]{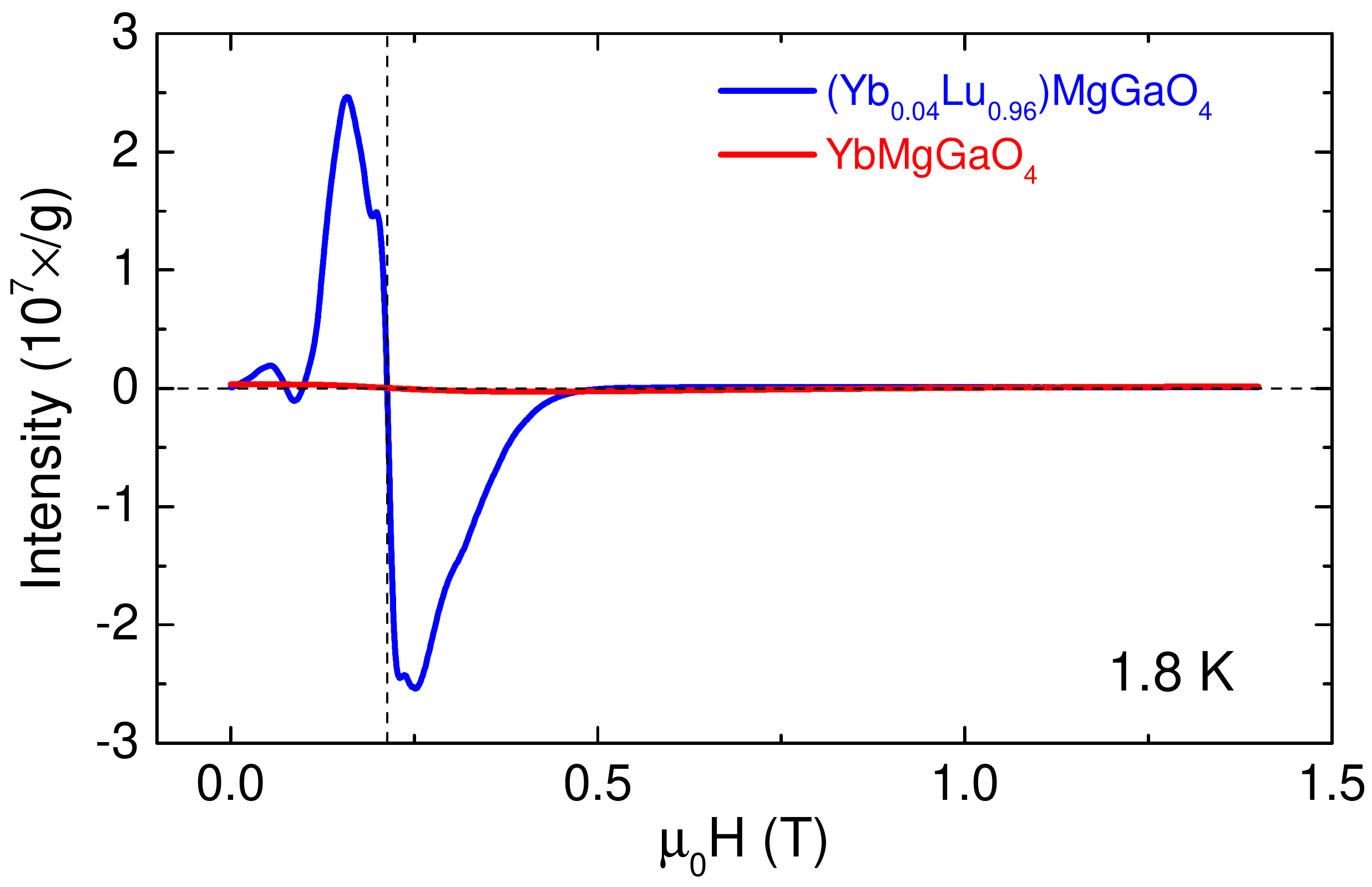}
\caption{(Color online.)
First-derivative absorption ESR spectra for (Yb$_{0.04}$Lu$_{0.96}$)MgGaO$_4$ and YbMgGaO$_4$ measured at 1.8 K, and normalized by the weight~\cite{li2015gapless}.}
\label{figs2}
\end{figure}

If a fraction of the magnetic Yb$^{3+}$ ions occupies the Mg$^{2+}$/Ga$^{3+}$ site, quasi-free spins of the Yb$^{3+}$ defects should be detected by ESR. The narrow and strong hyperfine lines~\cite{misra1998epr} from quasi-free defect Yb$^{3+}$ spins observed in (Yb$_{0.04}$Lu$_{0.96}$)MgGaO$_4$ completely disappear in YbMgGaO$_4$ (see Fig.~\ref{figs2}). From the comparison of the normalized signal intensities, we can estimate that the concentration of the Yb$^{3+}$-related defects in YbMgGaO$_4$ should be less than $\sim$ 0.04\%.

\section{4. Shape of the 97 meV CEF excitation}

The zoom-in view of the 97\,meV CEF excitation is shown in Fig. 2 (d) of the main text. It has a local maximum of the intensity around 97 meV, and the one-peak fit to this CEF excitation gives an averaged peak center at 96.6(2) meV (see fit \#1). However, the energy profile of this excitation is strongly asymmetric, with a clear shoulder at $\sim$ 87 meV.

\begin{figure}[H]
\centering
\includegraphics[width=10cm,angle=0]{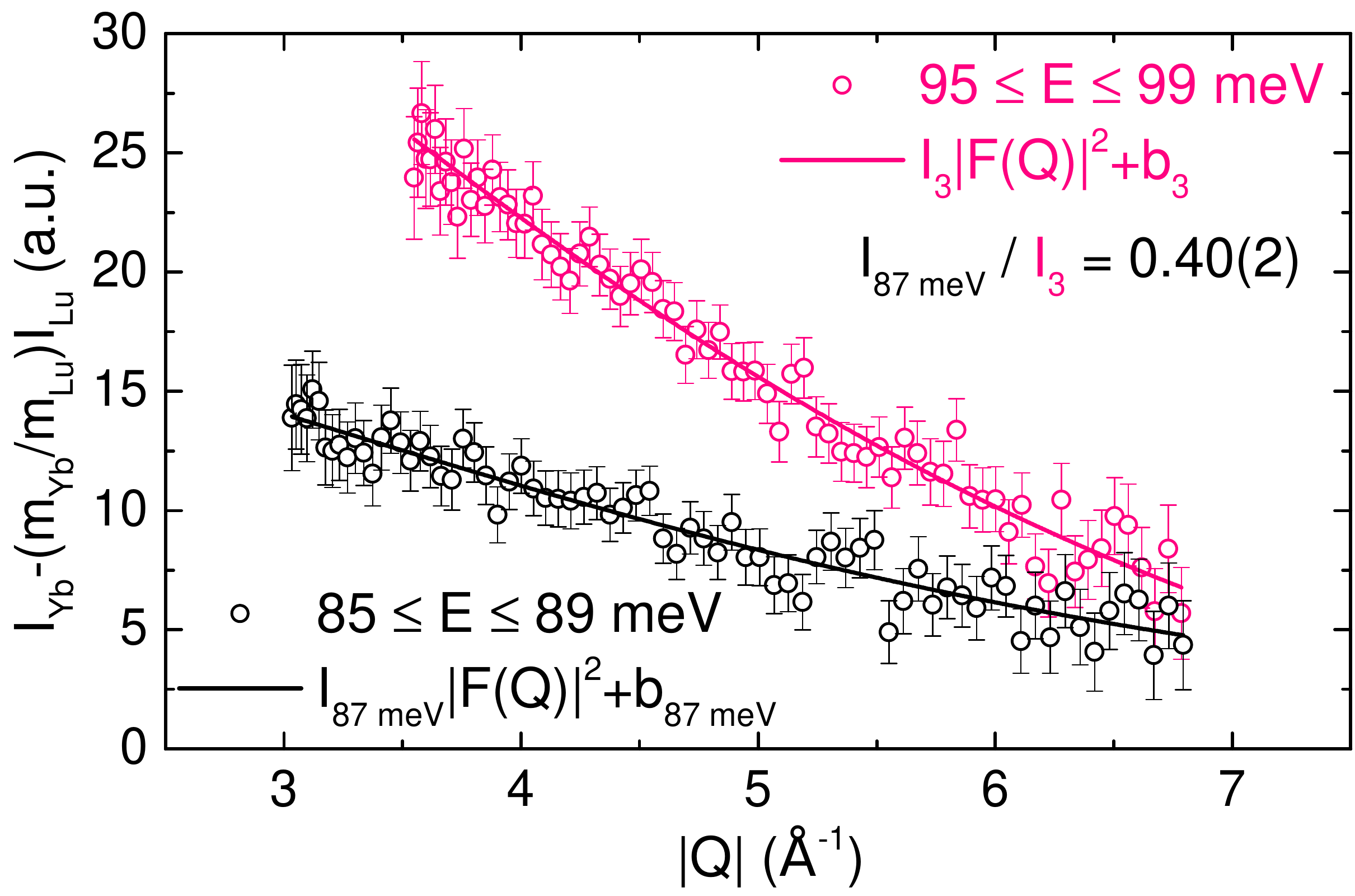}
\caption{(Color online.)
Momentum transfer ($\mid$Q$\mid$) dependences of the INS intensities integrated over the energy ranges, 85 $\leq$ E $\leq$ 89 meV and 95 $\leq$ E $\leq$ 99 meV, respectively.}
\label{figs3}
\end{figure}

The momentum transfer ($\mid$Q$\mid$) dependence of the integrated INS intensities at $\sim$ 87 meV and $\sim$ 97 meV are shown in Fig.~\ref{figs3}. Both intensities follow the magnetic form factor of the free Yb$^{3+}$ very well (see Fig.~\ref{figs3}), suggesting the same CEF origin of these spectral features.

The INS intensity of the shoulder at $\sim$ 87 meV is about 40\% of that around 97\,meV (see Fig.~\ref{figs3}). Neither Yb$^{3+}$-related impurities (such as Yb$_3$Ga$_5$O$_{12}$ and Yb$_2$O$_3$, $<$ 5\%) nor defects (such as the site mixing between Yb$^{3+}$ and Mg$^{2+}$/Ga$^{3+}$, $<$ 0.04\%) can account for this shoulder at $\sim$ 87 meV. Thus, the $\sim$ 87 meV shoulder is intrinsic and reflects a peculiar distribution of the CEF parameters in YbMgGaO$_4$.

\section{5. Energy dependence of the CEF excitations measured for YbMgGaO$_4$ from 5 to 300 K}

\begin{figure}[H]
\begin{center}
\includegraphics[width=10cm,angle=0]{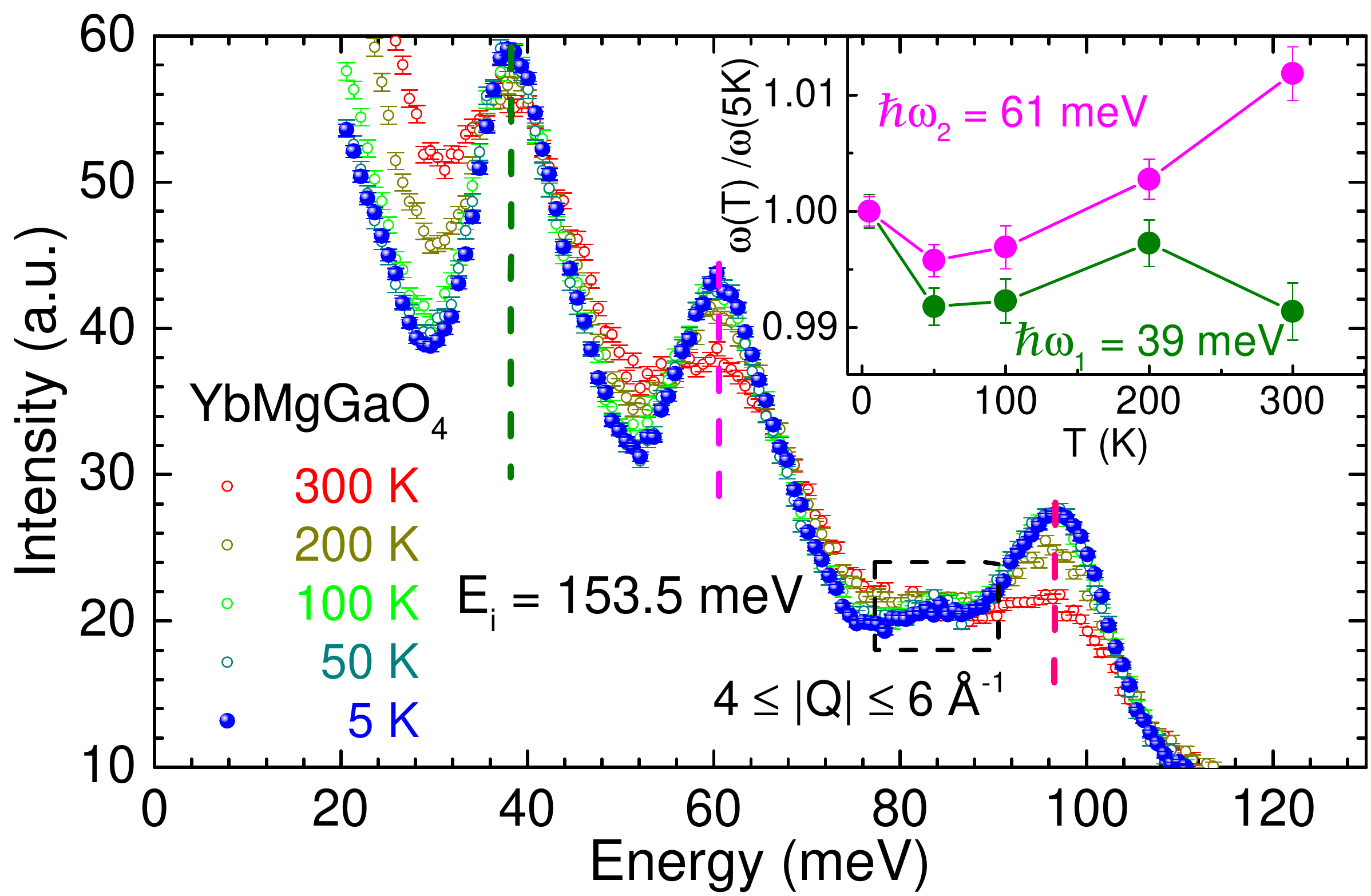}
\caption{(Color online)
Energy dependence of the INS intensity measured with the incident neutron energy of 153.5\,meV for YbMgGaO$_4$ at different temperatures. The data have been integrated over the same wave-vector space, $4\leq |Q|\leq 6$\,\r A$^{-1}$. Three CEF excitations of Yb$^{3+}$ are highlighted by vertical colored dashed lines. The black dashed rectangle shows the shoulder around 87\,meV. The inset shows temperature dependence of the first two excitation energies, $\hbar \omega_1(T)/\hbar \omega_1(\rm{5\,K})$ and $\hbar \omega_2(T)/\hbar \omega_2(\rm{5\,K})$.}
\label{figs4}
\end{center}
\end{figure}

No systematic anharmonic effect is observed excluding phonon origin of the excitations at $\hbar \omega$ = 39, 61, and 97\,meV (see Fig.~\ref{figs4}) ~\cite{PhysRevB.28.1928}.

\section{6. Combined CEF fit results for YbMgGaO$_4$}

\begin{table}[H]
\caption{Combined CEF fits for YbMgGaO$_4$, as described in the main text. Fit \#1 is performed against the whole dataset. For the fit \#2, the region between 73 and 90\,meV is excluded. For the fit \#3, the energy of the 87\,meV shoulder is used as the third CEF excitation. For the fit \#4, the energy of the 139\,meV mode is used as the third CEF excitation, see text for further details.}\label{table1}
\begin{center}
\begin{tabular}{ l || l | l || l | l || l | l || l | l }
    \hline
    \hline
    fits & \multicolumn{2}{|l||}{\textbf{\#1}} & \multicolumn{2}{|l||}{\#2} & \multicolumn{2}{|l||}{\#3} & \multicolumn{2}{|l}{\#4} \\ \hline
      & $X_i^{\rm obs}$ & $X_i^{\rm cal}$ & $X_i^{\rm obs}$ & $X_i^{\rm cal}$ & $X_i^{\rm obs}$ & $X_i^{\rm cal}$ & $X_i^{\rm obs}$ & $X_i^{\rm cal}$ \\ \hline
    $\hbar\omega_1$(meV) & 39.36(8) & 39.36 & 39.31(8) & 39.31 & 39.37(7) & 39.24 & 39.36(8) & 39.50\\
    $\hbar\omega_2$(meV) & 61.3(1) & 61.3 & 61.3(1) & 61.1 & 61.3(1) & 60.8 & 61.3(1) & 62.18\\
    $\hbar\omega_3$(meV) & 96.6(2) & 96.4 & 97.0(2) & 97.1 & 87(1) & 89 & 138.7(3) & 136.1\\
    $I_2/I_1$ & 0.72(2) & 0.70 & 0.72(2) & 0.78 & 0.72(2) & 0.84 & 0.72(2) & 0.41\\
    $I_3/I_1$ & 0.71(2) & 0.91 & 0.71(2) & 0.82 & 0.71(2) & 0.69 & 0.71(2) & 0.79\\
    $g_{\perp}$~\cite{li2015rare} & 3.06(2) & 3.21 & 3.06(2) & 3.19 & 3.06(2) & 3.18 & 3.06(2) & 3.02\\
    $g_{\parallel}$~\cite{li2015rare} & 3.72(3) & 3.68 & 3.72(3) & 3.67 & 3.72(3) & 3.44 & 3.72(3) & 4.06\\
    \hline
    $B_{2}^{0}$(meV) & \multicolumn{2}{|l||}{$-0.130$} & \multicolumn{2}{|l||}{$-0.141$} & \multicolumn{2}{|l||}{$-0.108$} & \multicolumn{2}{|l}{$0.571$} \\
    $B_{4}^{0}$(meV) & \multicolumn{2}{|l||}{$0.00179$} & \multicolumn{2}{|l||}{$0.0020$} & \multicolumn{2}{|l||}{$-0.0013$} & \multicolumn{2}{|l}{$0.0091$} \\
    $B_{4}^{3}$(meV) & \multicolumn{2}{|l||}{$-0.93$} & \multicolumn{2}{|l||}{$-0.96$} & \multicolumn{2}{|l||}{$-0.92$} & \multicolumn{2}{|l}{$-1.34$} \\
    $B_{6}^{0}$(meV) & \multicolumn{2}{|l||}{$-0.00095$} & \multicolumn{2}{|l||}{$-0.00096$} & \multicolumn{2}{|l||}{$-0.00082$} & \multicolumn{2}{|l}{$-0.0012$} \\
    $B_{6}^{3}$(meV) & \multicolumn{2}{|l||}{$-0.0247$} & \multicolumn{2}{|l||}{$-0.0239$} & \multicolumn{2}{|l||}{$-0.0186$} & \multicolumn{2}{|l}{$-0.0455$} \\
    $B_{6}^{6}$(meV) & \multicolumn{2}{|l||}{$0.0190$} & \multicolumn{2}{|l||}{$0.0169$} & \multicolumn{2}{|l||}{$0.0151$} & \multicolumn{2}{|l}{$0.00179$} \\
    \hline
    \hline
\end{tabular}
\end{center}
\end{table}

\section{7. Fitted CEF wavefunctions}

\begin{table}[H]
\caption{Fitted CEF wavefunctions.}\label{table2}
\begin{center}
\begin{tabular}{ l }
    \hline
    \hline
Fit \textbf{\#1} : \\
    \hline
$|\omega_{0,\pm}\rangle$ = $\pm0.71|\pm7/2\rangle\mp0.36|\mp5/2\rangle+0.60|\pm1/2\rangle$ \\
$|\omega_{1,\pm}\rangle$ = $0.97|\pm3/2\rangle\pm0.24|\mp3/2\rangle$ \\
$|\omega_{2,\pm}\rangle$ = $\mp0.21|\pm7/2\rangle+0.02|\mp7/2\rangle+0.07|\pm5/2\rangle\mp0.92|\mp5/2\rangle-0.31|\pm1/2\rangle\mp0.02|\mp1/2\rangle$ \\
$|\omega_{3,\pm}\rangle$ = $0.67|\pm7/2\rangle+0.09|\mp5/2\rangle\mp0.74|\pm1/2\rangle$ \\
    \hline
Fit \#2 : \\
    \hline
$|\omega_{0,\pm}\rangle$ = $\pm0.71|\pm7/2\rangle\mp0.35|\mp5/2\rangle+0.61|\pm1/2\rangle$ \\
$|\omega_{1,\pm}\rangle$ = $|\pm3/2\rangle$ \\
$|\omega_{2,\pm}\rangle$ = $\mp0.25|\pm7/2\rangle-0.02|\mp7/2\rangle-0.09|\pm5/2\rangle\mp0.93|\mp5/2\rangle-0.25|\pm1/2\rangle\pm0.02|\mp1/2\rangle$ \\
$|\omega_{3,\pm}\rangle$ = $0.66|\pm7/2\rangle+0.03|\mp5/2\rangle\mp0.75|\pm1/2\rangle$ \\
    \hline
Fit \#3 : \\
    \hline
$|\omega_{0,\pm}\rangle$ = $\pm0.70|\pm7/2\rangle\mp0.35|\mp5/2\rangle+0.62|\pm1/2\rangle$ \\
$|\omega_{1,\pm}\rangle$ = $0.71|\pm3/2\rangle\mp0.71|\mp3/2\rangle$ \\
$|\omega_{2,\pm}\rangle$ = $\mp0.29|\pm7/2\rangle+0.03|\mp7/2\rangle+0.11|\pm5/2\rangle\mp0.93|\mp5/2\rangle-0.21|\pm1/2\rangle\mp0.02|\mp1/2\rangle$ \\
$|\omega_{3,\pm}\rangle$ = $0.66|\pm7/2\rangle-0.03|\mp5/2\rangle\mp0.75|\pm1/2\rangle$ \\
    \hline
Fit \#4 : \\
    \hline
$|\omega_{0,\pm}\rangle$ = $\pm0.66|\pm7/2\rangle+0.08|\mp7/2\rangle-0.02|\pm5/2\rangle\mp0.14|\mp5/2\rangle+0.73|\pm1/2\rangle\mp0.09|\mp1/2\rangle$ \\
$|\omega_{1,\pm}\rangle$ = $|\pm3/2\rangle$ \\
$|\omega_{2,\pm}\rangle$ = $\mp0.15|\pm7/2\rangle-0.03|\pm5/2\rangle\mp0.99|\mp5/2\rangle-0.05|\pm1/2\rangle$ \\
$|\omega_{3,\pm}\rangle$ = $0.73|\pm7/2\rangle-0.08|\mp5/2\rangle\mp0.68|\pm1/2\rangle$ \\
    \hline
    \hline
\end{tabular}
\end{center}
\end{table}

\section{8. Crystal structure of YbMgGaO$_4$ at 100 K determined from single-crystal XRD data}

In Table~\ref{table3}, we provide average crystal structure of YbMgGaO$_4$ determined from single-crystal XRD data at 100\,K (the data are from the Ref.~\cite{li2015rare}). When Yb$^{3+}$ is placed into its ideal position at $(0,0,0)$, the displacement parameter U$_{33}$ is nearly twice larger than U$_{11}$ = U$_{22}$. This reflects static disorder of Yb$^{3+}$ along the \emph{c}-axis~\cite{cava1998compounds}, because the large U$_{33}$ of 0.0240 {\AA}$^2$ persists even at 100 K, where dynamic effects, such as phonons, should be minor. Mg$^{2+}$ and Ga$^{3+}$ are intermixed. On the other hand, no evidence for Yb antisite defects (for example, Yb atoms occupying the Mg/Ga position) is found.

\begin{table}[H]
\centering
\caption{Crystal structure of YbMgGaO$_4$ at 100 K determined from single-crystal XRD data (\emph{a} = 3.4061 {\AA} and \emph{c} = 25.1297 {\AA}). In contrast to Ref.~\cite{li2015rare}, we considered the fully position ordered model with the Yb atoms at (0,0,0), resulting in the elongated thermal ellipsoid for Yb. Alternatively, this elongation can be described by the split position of Yb, thus producing the disordered structural model reported in Ref.~\cite{li2015rare}. However, our current analysis invalidates such a model, because the Yb atoms are not simply split into two positions, as the model would suggest, and rather show a continuous spread along the c-axis, following the disordered arrangement of Mg and Ga in the structure.}
\label{table3}
\begin{tabular}{c  c  c}
 & & \\
 & & \\
\hline
\hline
\multicolumn{3}{c}{Space group: $R\bar{3}m$} \\
\hline
Yb & z & 0 \\
   & U$_{11}$ = U$_{22}$ = 2U$_{12}$ & 0.0127(3) \\
x = y = 0 & U$_{33}$ & 0.0240(4) \\
   & \textbf{U$_{33}$/U$_{11}$} & \textbf{1.89(7)} \\
U$_{13}$ = U$_{23}$ = 0 & Occupancy & 1 \\
\hline
Mg/Ga & z & 0.21442(7) \\
   & U$_{11}$ = U$_{22}$ = 2U$_{12}$ & 0.0184(6) \\
x = y = 0 & U$_{33}$ & 0.0134(8) \\
   & \textbf{U$_{33}$/U$_{11}$} & \textbf{0.73(7)} \\
U$_{13}$ = U$_{23}$ = 0 & Occupancy & 0.5 \\
\hline
O1 & z & 0.2912(3) \\
   & U$_{11}$ = U$_{22}$ = 2U$_{12}$ & 0.020(2) \\
x = y = 0 & U$_{33}$ & 0.015(4) \\
   & \textbf{U$_{33}$/U$_{11}$} & \textbf{0.8(3)} \\
U$_{13}$ = U$_{23}$ = 0 & Occupancy & 1 \\
\hline
O2 & z & 0.1284(4) \\
   & U$_{11}$ = U$_{22}$ = 2U$_{12}$ & 0.027(3) \\
x = y = 0 & U$_{33}$ & 0.018(4) \\
   & \textbf{U$_{33}$/U$_{11}$} & \textbf{0.7(2)} \\
U$_{13}$ = U$_{23}$ = 0 & Occupancy & 1 \\
\hline
\multicolumn{2}{c}{Number of variables} & 12 \\
\hline
Residuals & R(F) (I $>$ 3$\sigma_I$) & 0.049 \\
          & R$_w$(F) (I $>$ 3$\sigma_I$) & 0.063 \\
\hline
\hline
\end{tabular}
\end{table}

\section{9. Local structural configurations of Yb$^{3+}$ obtained from density-functional band-structure calculations}

In order to go beyond the average structure from single-crystal XRD, we constructed several ordered Mg/Ga configurations and optimized their geometry using density-functional (DFT) band-structure calculations. All calculations were performed in the \texttt{VASP} code~\cite{vasp1,vasp2} with the Perdew-Burke-Ernzerhof (PBE) flavor of the exchange-correlation potential~\cite{pbe96} and the default Yb pseudopotential, where $4f$ states were placed into the core. The $4\times 4\times 4$ k-mesh was used, and residual forces were below 0.005\,eV/\r A in fully optimized structures. Lattice parameters were fixed to their experimental values.

\begin{figure}[H]
\centering
\includegraphics[width=10cm,angle=0]{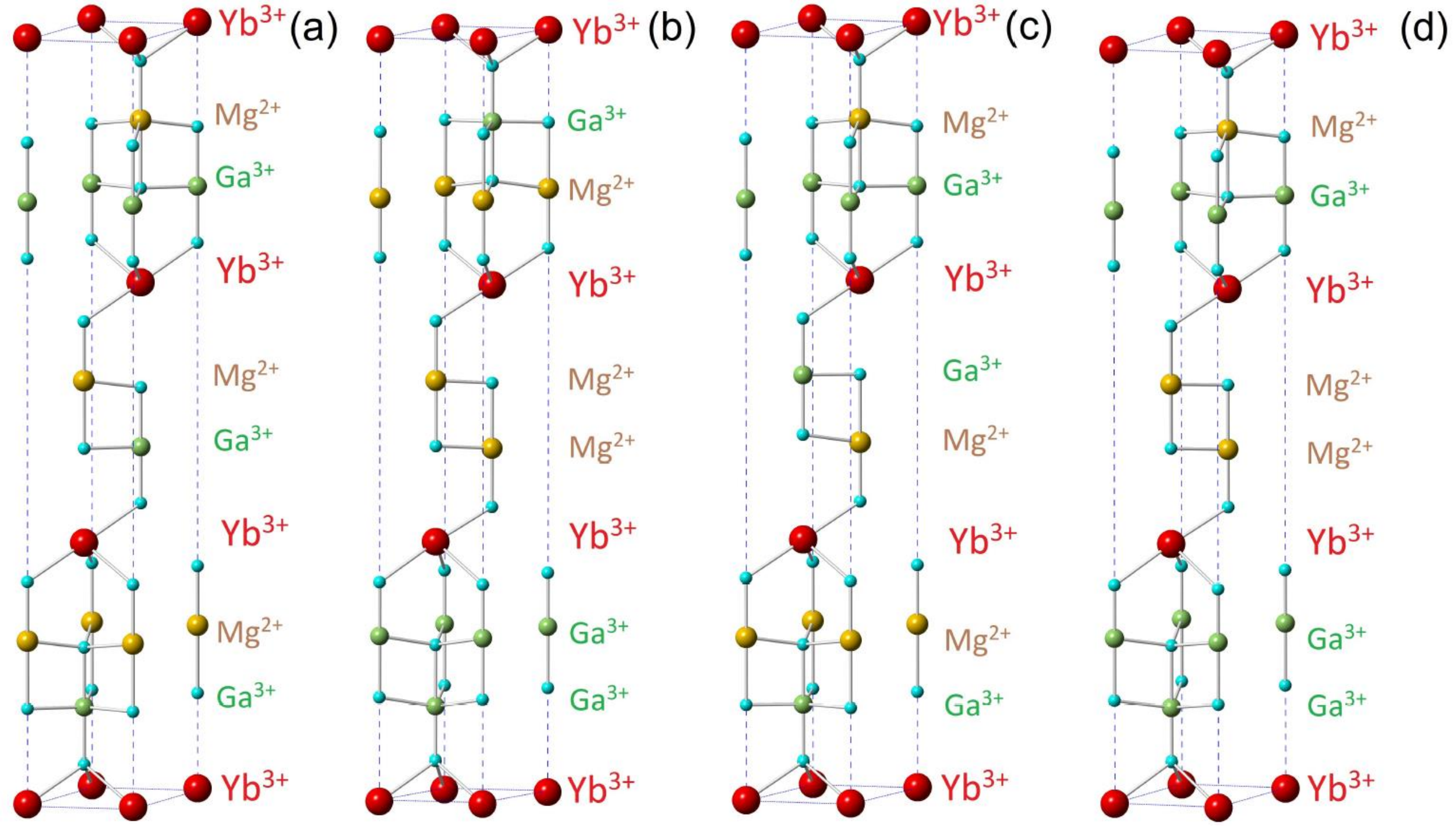}
\caption{(Color online.)
Optimized crystal structures for charge-neutral unit cells containing 7 possible environments of Yb$^{3+}$: (a) YbMgGaYbMgGaYbMgGaYb is used to extract the local structure of the YbO$_6$ octahedron 2 (see Table~\ref{table4}), (b) YbGaMgYbMgMgYbGaGaYb is used to extract the local structures of the YbO$_6$ octahedra 3, 4 and 7 (see Table~\ref{table4}), (c) YbMgGaYbGaMgYbMgGaYb is used to extract the local structures of the YbO$_6$ octahedra 1 and 2 (see Table~\ref{table4}), and (d) YbMgGaYbMgMgYbGaGaYb is used to extract the local structures of the YbO$_6$ octahedra 5, 6 and 7 (see Table~\ref{table4}).}
\label{figs5}
\end{figure}

The local configurations considered in our analysis are periodic and charge-neutral. Additionally, we restricted ourselves to those configurations that preserve the 3-fold symmetry of the Yb$^{3+}$ site. While many other configurations violating this symmetry can be constructed, the 3-fold-symmetric configurations keep the problem tractable and are in fact sufficient to reproduce effects of CEF randomness in YbMgGaO$_4$. We considered four different Mg/Ga distributions within the trigonal unit cell of YbMgGaO$_4$ (Fig.~\ref{figs5}). These four structures contain all 7 possible environments of Yb, where we define the environment with respect to the Mg/Ga distribution in the two neighboring slabs. This way, 7 A-B-Yb-C-D configurations, where each of the ABCD is either Mg or Ga, are obtained (Table~\ref{table4}). The restriction to only two neighboring slabs is well justified. For example, the MgMgYbGaGa configuration is independently probed in the structures shown in panels (b) and (d) of Fig.~\ref{figs5} resulting in less than 0.5\,\% deviation for the ensuing geometry of the YbO$_6$ octahedron.

We find that the exact position of Yb as well as its oxygen coordination are strongly affected by the Mg/Ga distribution. The local configurations MgGaYbGaMg and GaMgYbMgGa retain inversion symmetry at the Yb site and keep the YbO$_6$ octahedron undistorted. Any other configuration entails electrostatic field that pushes Yb atoms out of their ideal $(0,0,0)$ position. This effect is also detected experimentally by the largely elongated thermal ellipsoid in the averaged crystal structure of YbMgGaO$_4$ (Table~\ref{table3}). Consequently, the YbO$_6$ octahedra are distorted, as shown in Table~\ref{table4}.

\begin{table}[H]
\centering
\caption{Optimized geometries of the YbO$_6$ octahedra for different stacking sequences of Mg/Ga. Here, $R_{O1}$ is the distance between Yb and its neighboring O1, and $\theta_{\rm O1}$ is the angle between the Yb--O1 bond and the \emph{c}-axis. For undistorted YbO$_6$ octahedra, there is only one set of $R_{\rm O1}$ and $\theta_{\rm O1}$, whereas two sets of $R_{\rm O1}$ and $\theta_{\rm O1}$ are obtained for the distorted YbO$_6$ octahedra.}
\label{table4}
\begin{tabular}{c  c  c  c  c}
 & & & &\\
 & & & &\\
\hline
\hline
YbO$_6$ octahedrons & Local stackings & Probability & $R_{O1}$ & $\theta_{O1}$ \\
\hline
1 & MgGaYbGaMg/ & 1/7 & 2.2615 & 60.3953 \\
  & GaMgYbMgGa  &     & undistorted	& undistorted \\
\hline
2 & MgGaYbMgGa/ & 1/7 & 2.2685 & 60.1025 \\
  & GaMgYbGaMg  &     & 2.2570 & 60.6052 \\
\hline
3 & GaMgYbMgMg/ & 1/7 & 2.260 & 60.456 \\
  & MgMgYbMgGa  &     & 2.251 & 60.865 \\
\hline
4 & GaGaYbGaMg/ & 1/7 & 2.321 & 57.927 \\
  & MgGaYbGaGa  &     & 2.243 & 61.244 \\
\hline
5 & MgGaYbMgMg/ & 1/7 & 2.258 & 60.552 \\
  & MgMgYbGaMg  &     & 2.240 & 61.393 \\
\hline
6 & GaGaYbMgGa/ & 1/7 & 2.316 & 58.105 \\
  & GaMgYbGaGa  &     & 2.247 & 61.054\\
\hline
7 & GaGaYbMgMg/ & 1/7 & 2.321 & 57.904 \\
  & MgMgYbGaGa  &     & 2.228 & 61.972 \\
\hline
\hline
\end{tabular}
\end{table}

\section{10. Calculated CEF randomness in the framework of the point-charge model}

\subsection{10.1 Calculated CEF randomness caused by the distribution of nearest-neighbor oxygen environments}

\begin{figure}[H]
\centering
\includegraphics[width=10cm,angle=0]{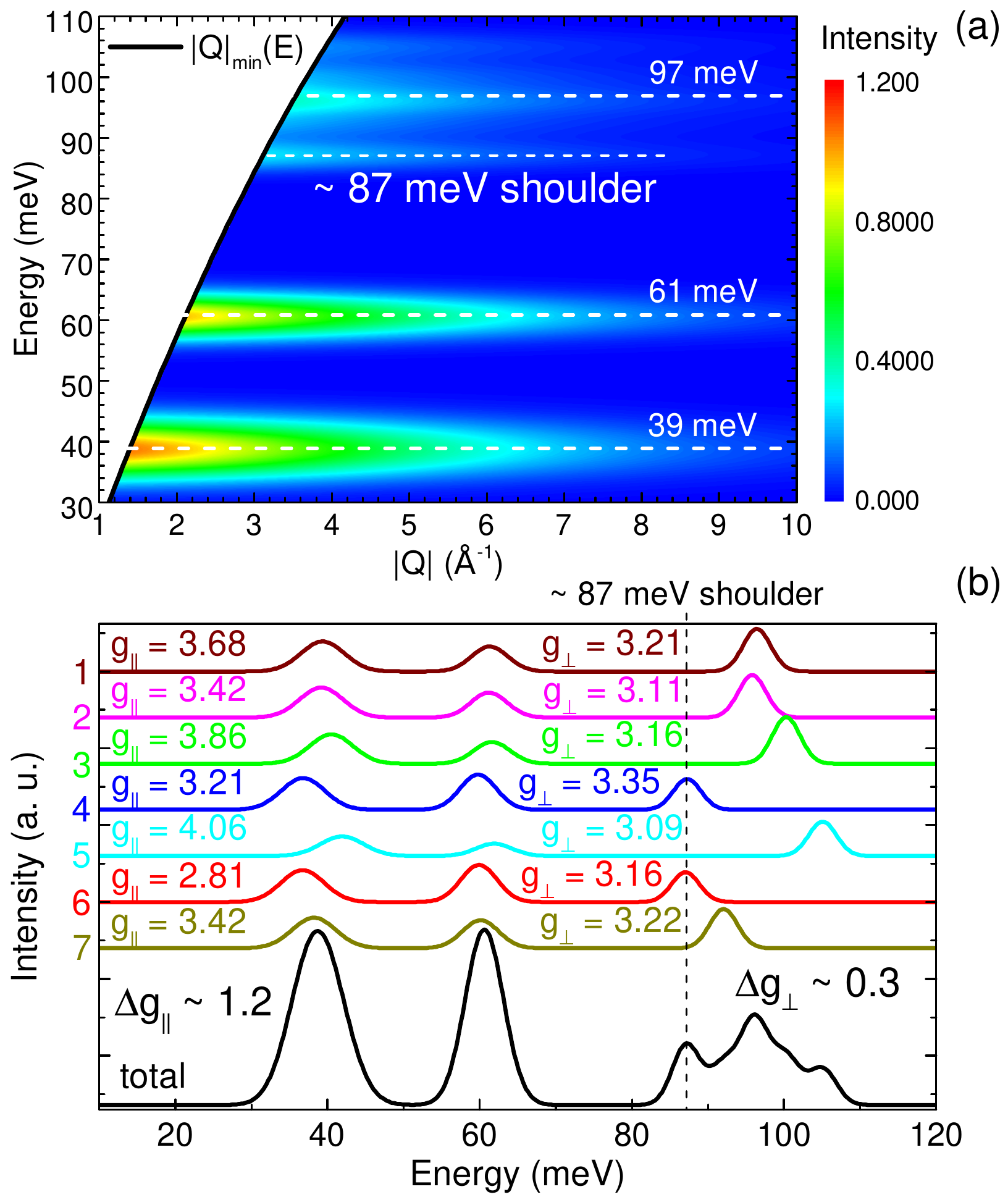}
\caption{(Color online.)
(a) INS spectra calculated using the distribution of the CEF parameters from Table~\ref{table5} with $\langle B_n^m\rangle$ = $B_n^m$(fit \#1), convoluted with the corresponding instrumental resolutions. The black line represents the measurable $|Q|_{\min}$. (b) Calculated energy dependences of the CEF INS intensity using the CEF parameter series for different YbO$_6$ octahedra (see Table~\ref{table5}). The black curve shows the cumulative energy profile of the CEF excitations. The resulting effective spin-1/2 $g$-factors for each local configuration are shown. The variable distortions of the YbO$_6$ octahedra introduce the overall distribution of the $g$-values by $\Delta g_{\parallel}\sim 1.2$ and $\Delta g_{\perp}\sim 0.3$.}
\label{figs6}
\end{figure}

In Table~\ref{table5}, we list CEF parameters calculated using the point-charge model for 7 flavors of the YbO$_6$ octahedra listed in Table~\ref{table4}. Relative distributions listed in Table~\ref{table5} are then applied to obtain the CEF spectra by setting $\langle B_n^m\rangle$ = $B_n^m$(fit \#1) (see Table~\ref{table1} for each $B_n^m$(fit \#1)). Each type of the YbO$_6$ octahedron produces its three CEF excitations at slightly different energies (Fig.~\ref{figs6}). The cumulative effect of this distribution is the symmetric broadening of the two lower CEF excitations and the complex profile of the third (highest-energy) CEF excitation, where the configurations 4 and 6 are responsible for the shoulder on the low-energy side (around 87\,meV), whereas the configuration 5 would produce the shoulder on the high-energy side (around 105\,meV).

The resulting energy profile is remarkably similar to our experimental observations, as further demonstrated by the 2D plot in Fig. 3 (d) in the main text. We note, however, that the much weaker (about a half of the intensity of the $\sim$ 87\,meV shoulder) high-energy shoulder around 105\,meV was not observed at its expected position. On the other hand, we do observe a weak excitation at a higher energy of 139\,meV (see below), which may be related to the high-energy shoulder.

\begin{table}[H]
\centering
\caption{Relative CEF parameters, $B_n^m$(YbO$_6$)/$\langle B_n^m\rangle$, obtained from the point-charge model for different Yb-O geometries from Table~\ref{table4}.}
\label{table5}
\begin{tabular}{c | c | c | c | c | c | c | c}
\multicolumn{8}{c}{  }  \\
\multicolumn{8}{c}{  }  \\
\hline
\hline
YbO$_6$ octahedra & Probability & $n=2$; $m=0$ & $n=4$; $m=0$ & $n=4$; $m=3$ & $n=6$; $m=0$ & $n=6$; $m=3$ & $n=6$; $m=6$ \\
\hline
1 & 1/7 & 1.0000 & 1.0000 & 1.0000 & 1.0000 & 1.0000 & 1.0000 \\
2 & 1/7 & 0.9916 & 1.0005 & 0.9973 & 0.9951 & 0.9766 & 0.9942 \\
3 & 1/7 & 1.0529 & 0.9897 & 1.0130 & 1.0240 & 1.1605 & 1.0352 \\
4 & 1/7 & 0.8510 & 1.0071 & 0.9570 & 0.9153 & 0.6276 & 0.9138 \\
5 & 1/7 & 1.1154 & 0.9749 & 1.0271 & 1.0496 & 1.3554 & 1.0768 \\
6 & 1/7 & 0.8481 & 1.0116 & 0.9578 & 0.9171 & 0.6064 & 0.9106 \\
7 & 1/7 & 0.9233 & 0.9880 & 0.9731 & 0.9429 & 0.8615 & 0.9646 \\
\hline
\multicolumn{2}{c|}{Expected deviations} & 0.0792 & 0.0095 & 0.0221 & 0.0433 & 0.2063 & 0.0470 \\
\hline
\hline
\end{tabular}
\end{table}

\subsection{10.2 Calculated CEF randomness caused solely by the site-mixing between Mg$^{2+}$ and Ga$^{3+}$}

We have also used the point-charge model to calculate the CEF parameters for different Mg/Ga distributions, while keeping all atoms in their ideal positions (see Table~\ref{table3}), i.e., neglecting effects of local structural relaxation. The distribution of the CEF parameters (see Table~\ref{table6}) appears to be minor compared to Table~\ref{table5} discussed above. This minor distribution broadens the CEF excitations by $\Delta(\hbar\omega_1)$ = 0.27 meV, $\Delta(\hbar\omega_2)$ = 0.26 meV and $\Delta(\hbar\omega_3)$ = 0.39 meV, which is much smaller than the broadenings of several meV observed experimentally. Therefore, the distribution of Mg/Ga does not explain the CEF randomness, and effects of local structural relaxation are crucial.

\begin{table}[H]
\centering
\caption{Relative CEF parameters, $B_n^m$(Mg/Ga)/$\langle B_n^m\rangle$, obtained from the point-charge model for different distributions of Mg$^{2+}$ and Ga$^{3+}$ around Yb$^{3+}$ (total 6 nearest-neighbor Mg/Ga sites around each Yb$^{3+}$). All ions are kept at their ideal positions obtained from XRD (see Table~\ref{table3}).}
\label{table6}
\begin{tabular}{c | c | c | c | c | c | c | c}
\multicolumn{8}{c}{  }  \\
\multicolumn{8}{c}{  }  \\
\hline
\hline
Mg/Ga configurations & Probability & $n=2$; $m=0$ & $n=4$; $m=0$ & $n=4$; $m=3$ & $n=6$; $m=0$ & $n=6$; $m=3$ & $n=6$; $m=6$ \\
\hline
6Mg & 1/64 & 0.8990 & 1.0113 & 1.0107 & 0.9891 & 0.9702 & 1.0005 \\
5Mg+Ga & 3/32 & 0.9327 & 1.0075 & 1.0071 & 0.9927 & 0.9801 & 1.0004 \\
4Mg+2Ga & 15/64 & 0.9663 & 1.0038 & 1.0036 & 0.9964 & 0.9901 & 1.0002 \\
3Mg+3Ga & 5/16 & 1.0000 & 1.0000 & 1.0000 & 1.0000 & 1.0000 & 1.0000 \\
2Mg+4Ga & 15/64 & 1.0337 & 0.9962 & 0.9964 & 1.0036 & 1.0099 & 0.9998 \\
Mg+5Ga & 3/32 & 1.0673 & 0.9925 & 0.9929 & 1.0073 & 1.0199 & 0.9996 \\
6Ga & 1/64 & 1.1010 & 0.9887 & 0.9893 & 1.0109 & 1.0298 & 0.9995 \\
\hline
\multicolumn{2}{c|}{Expected deviations} & 0.0315 & 0.0035 & 0.0033 & 0.0034 & 0.0093 & 0.00017 \\
\hline
\hline
\end{tabular}
\end{table}

\section{11. INS spectra measured at $E_i$ = 307\,meV: The 138.7(3) meV mode}

\begin{figure}[H]
\centering
\includegraphics[width=17cm,angle=0]{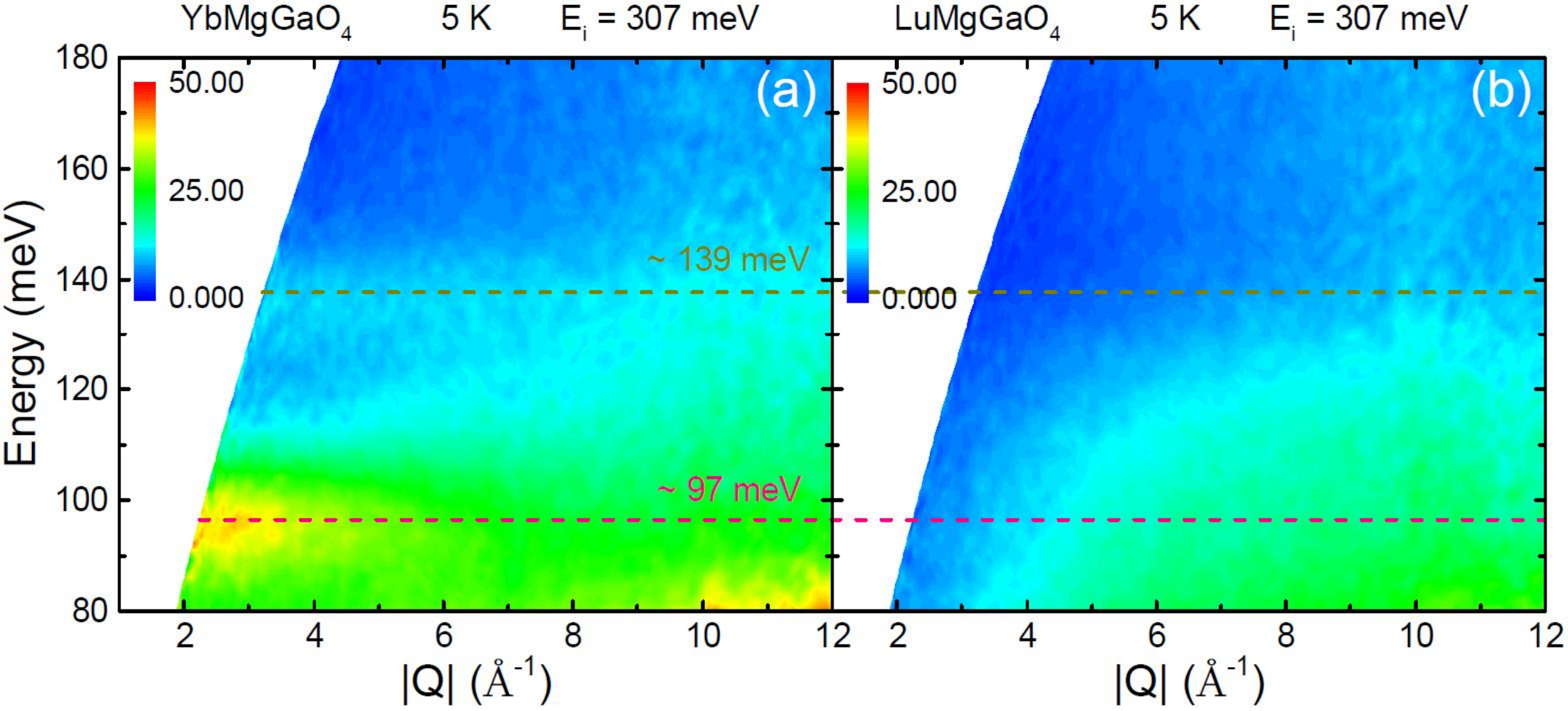}
\caption{(Color online.)
MERLIN INS spectra for (a) YbMgGaO$_4$ and (b) LuMgGaO$_4$ measured with the incident neutron energy of 307 meV at 5 K.}
\label{figr1}
\end{figure}

\begin{figure}[H]
\centering
\includegraphics[width=17cm,angle=0]{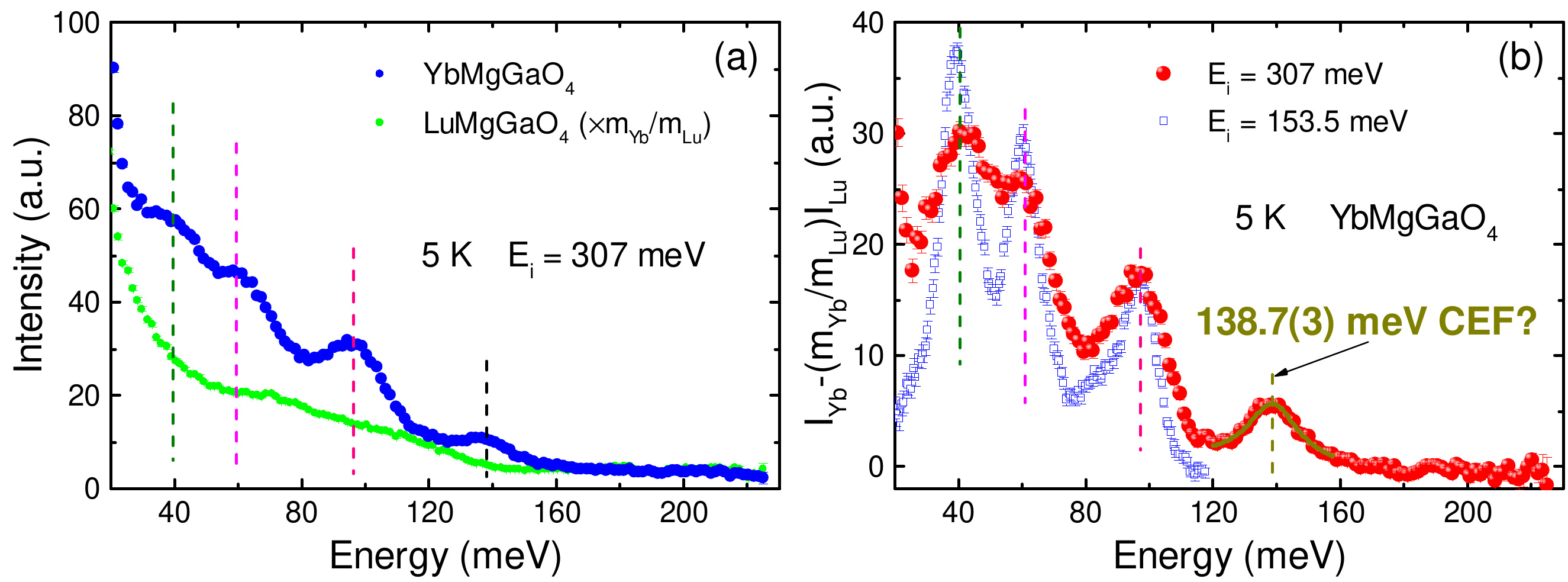}
\caption{(Color online.)
(a) Energy dependence of the INS intensity for YbMgGaO$_4$ and LuMgGaO$_4$ integrated over 4 $\leq |Q| \leq$ 6 {\AA}$^{-1}$ with the incident neutron energy of 307 meV at 5 K. (b) CEF excitations (4 $\leq |Q| \leq$ 6 {\AA}$^{-1}$) measured with the incident neutron energy of 307 and 153.5 meV at 5 K.}
\label{figr2}
\end{figure}

\begin{figure}[H]
\centering
\includegraphics[width=10cm,angle=0]{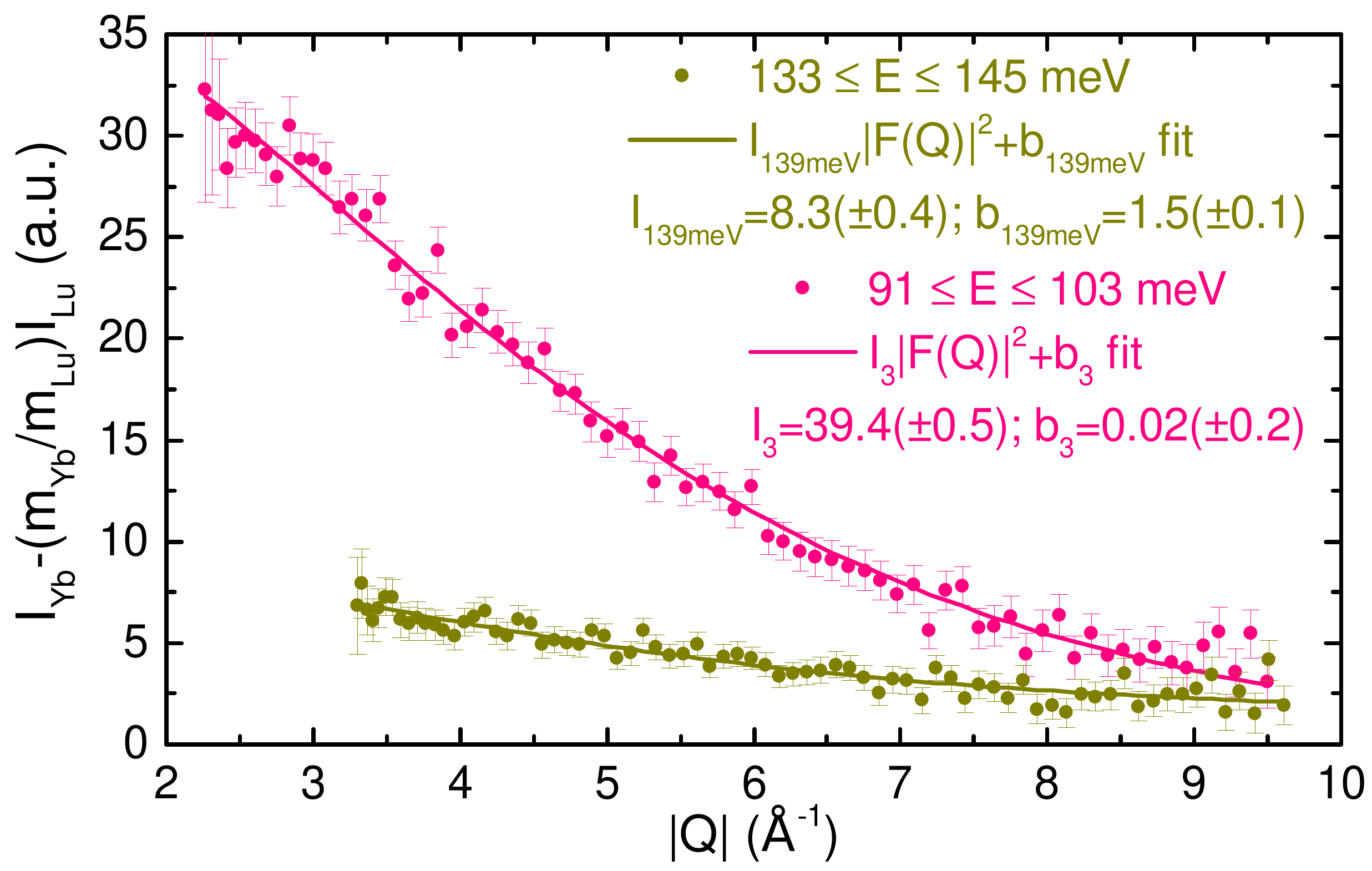}
\caption{(Color online.)
Wave-vector ($|Q|$) dependence of the INS intensity integrated over 91 $\leq E \leq$ 103 meV (pink) and 133 $\leq E \leq$ 145 meV (dark yellow), measured with $E_i$ = 307 meV at 5 K.}
\label{figr3}
\end{figure}

The incident energy of $E_i$ = 307 meV gives access to a much broader energy range and reveals a weak excitation around 139 meV (see Fig.~\ref{figr1}). After the subtraction by the corresponding INS signal of LuMgGaO$_4$, this $\sim$ 139 meV mode becomes more clear, and its position is fitted to be 138.7(3) meV (see Fig.~\ref{figr2}). Three main CEF excitations are confirmed, although they show significant broadening due to the poor instrumental resolution at $E_i$ = 307 meV ($\sigma$ $\sim$ 22.5 meV at $\hbar\omega$ $\sim$ 0 meV) (see Fig.~\ref{figr2} (b)). The intensity of the 138.7(3) meV mode is fitted to be $I_{139 meV}/I_3$ = 0.21(1) (see Fig.~\ref{figr3}), which is much smaller than that of the $\sim$ 87 meV shoulder, $I_{87 meV}/I_3$ = 0.40(2) (see Fig.~\ref{figs3}). Therefore, $I_{139 meV}/I_{87 meV}$ $\sim$ 0.5.

$|Q|$-dependence of the intensity of the 139\,meV mode confirms its CEF nature, although a sizable background of unknown origin can be seen from the non-zero value of $b$ in Fig.~\ref{figr3}. We assumed that this mode may be a trace of the third CEF excitations and used it for the CEF fit \#4, where the energy of the third excitation was set to 139\,meV (see Tables~\ref{table1} and~\ref{table2}). Owing to its low intensity, this mode may be the third CEF excitation for a small fraction of the Yb$^{3+}$ ions only (up to 20\,\%). Its intensity of one half of the 87\,meV shoulder corresponds in fact to the 105\,meV high-energy shoulder in our calculated spectra on Fig.~\ref{figs6}. An alternative and perhaps even more plausible explanation for the 139\,meV mode would be a multiple scattering CEF excitation, such as $\hbar\omega_1+\hbar\omega_3=136$\,meV and $2\hbar\omega_1+\hbar\omega_2=139$\,meV, because neutron can experience several scattering events before leaving the sample~\cite{PhysRevB.92.144422}.

We also note that our sequence of CEF excitations, including the weak 139\,meV mode, is well in line with the recent report by Paddison~\textit{et al.}~\cite{paddison2016continuous}.

\section{12. Elastic neutron diffraction pattern for the single crystal sample of YbMgGaO$_4$ in the \emph{ab}-plane used in LET INS experiments}

\begin{figure}[H]
\centering
\includegraphics[width=10cm,angle=0]{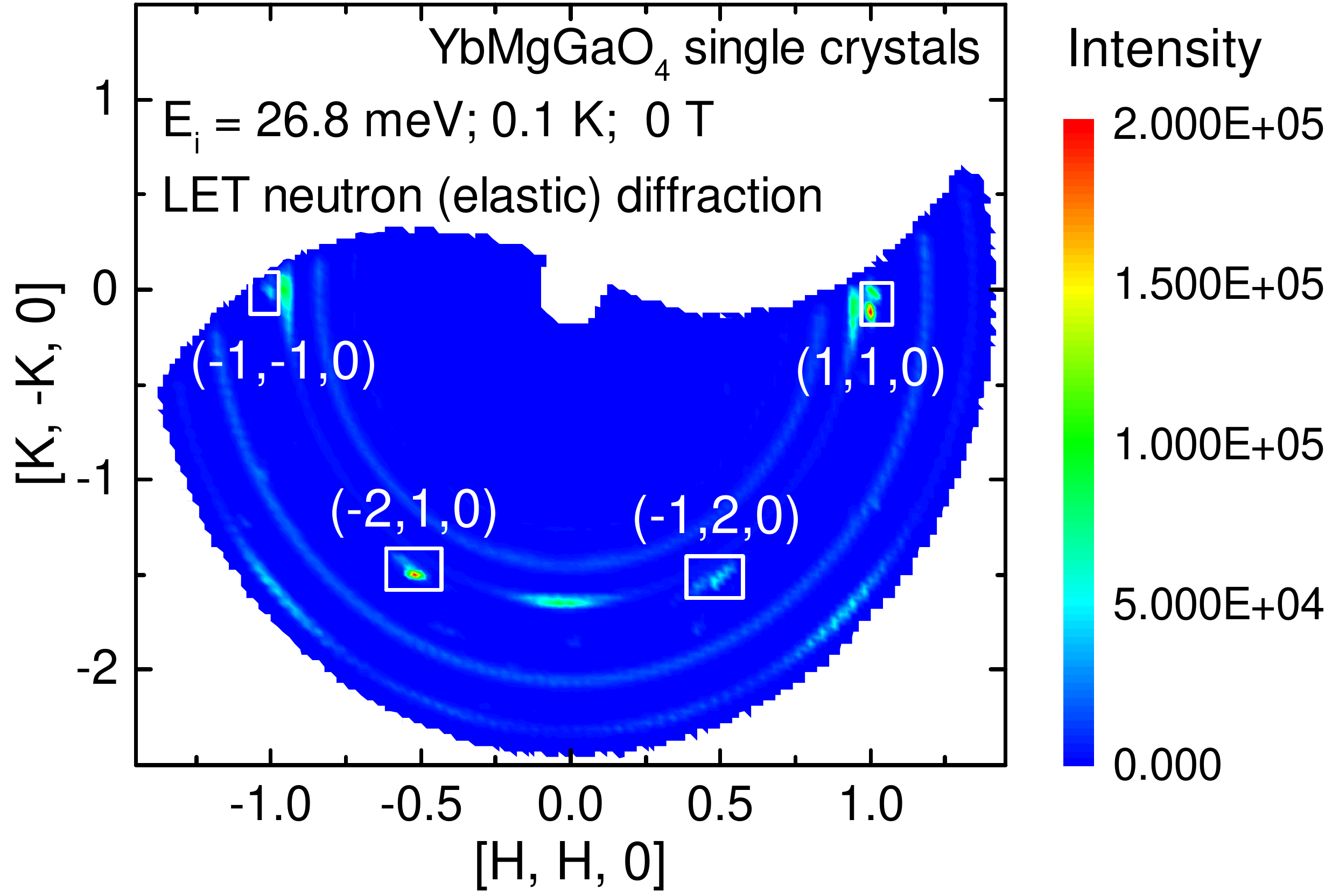}
\caption{(Color online.)
LET neutron (elastic) diffraction pattern for the YbMgGaO$_4$ single crystal sample at 0.1 K and 0 T.}
\label{figs8}
\end{figure}

\section{13. Magnetic field dependence of the susceptibility ($dM/dH$) measured at 1.9 K along the \emph{c}-axis}

\begin{figure}[H]
\centering
\includegraphics[width=10cm,angle=0]{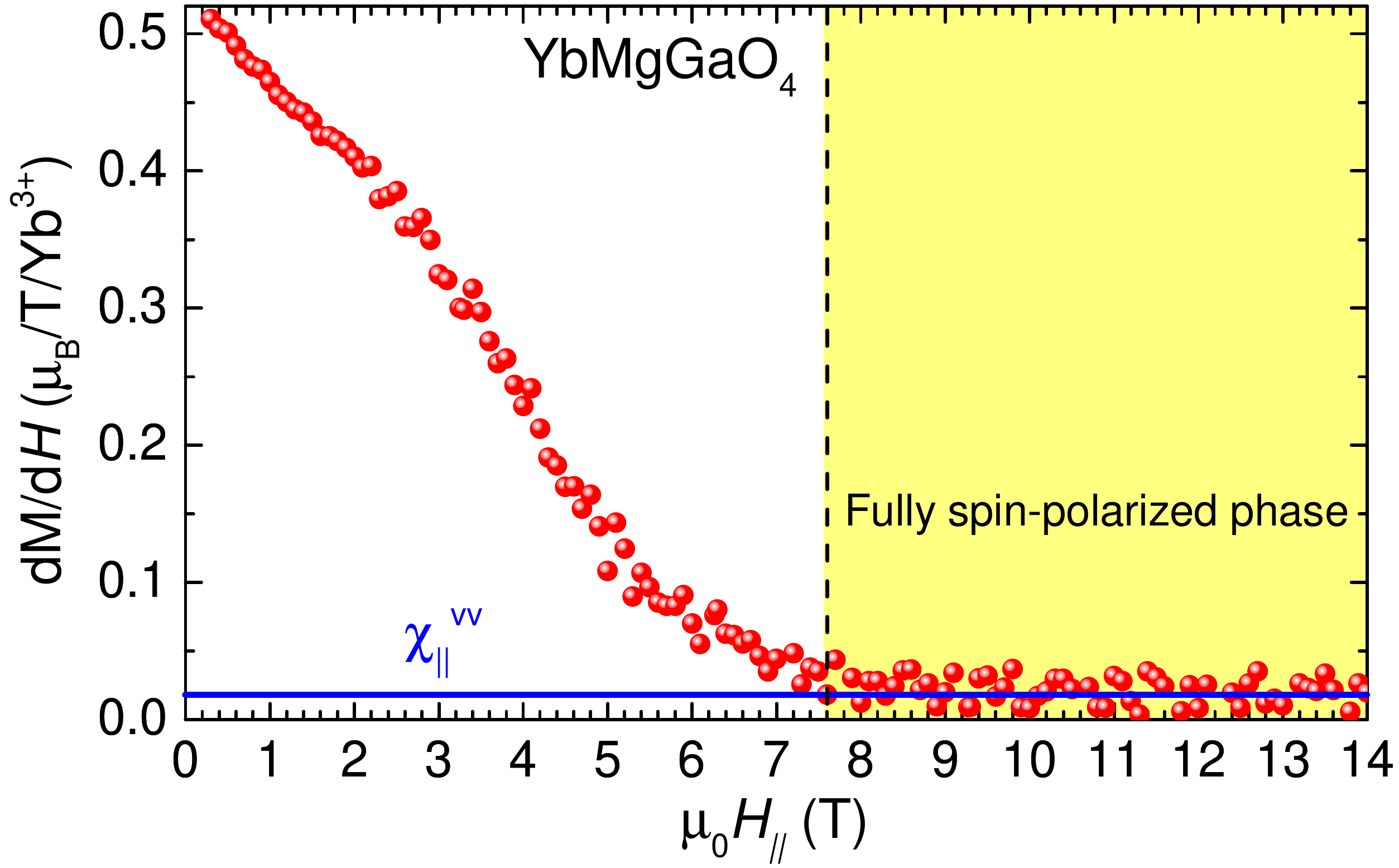}
\caption{(Color online.)
Magnetic field dependence of the susceptibility ($dM/dH$) measured at 1.9 K along the \emph{c}-axis. It
confirms full polarization of the material in the field of 8.5 T that was used for the LET measurement.}
\label{figs9}
\end{figure}

\bibliography{Merlin_YbMgGaO4}

\end{document}